\documentclass[prl, twocolumn, showpacs, reprint, nofootinbib, superscriptaddress, floatfix]{revtex4-2}
\usepackage[colorlinks = True, linkcolor = blue, citecolor = blue, breaklinks = True]{hyperref}
\usepackage{times}
\usepackage{xcolor}
\usepackage{graphicx}
\usepackage{subfigure}
\usepackage{setspace}
\usepackage{calc}
\usepackage{calrsfs}
\usepackage{float}
\usepackage[normalem]{ulem}
\usepackage{xcolor}
\usepackage{mathtools}
\usepackage{comment}
\usepackage{ulem}
\usepackage{amssymb,amsmath,amsfonts}
\usepackage{graphics,epsfig,subfigure}
\usepackage{color,bm,hyperref}
\usepackage{xcolor}
\usepackage{tikz}
\usepackage{yfonts}
\usetikzlibrary{shapes.geometric,arrows}
\usetikzlibrary{decorations.markings}
\DeclareMathAlphabet{\pazocal}{OMS}{zplm}{m}{n}

\begin{document}

\title{Manipulating phases in many-body interacting systems with subsystem resetting}
\author{Anish Acharya $^\mathsection$}
\email{Email: anish.acharya@tifr.res.in}
\affiliation{Department of Theoretical Physics, Tata Institute of Fundamental Research, Homi Bhabha Road, Mumbai 400005, India}
\author{Rupak Majumder $^\mathsection$}
\email{Email: rupak.majumder@tifr.res.in}
\affiliation{Department of Theoretical Physics, Tata Institute of Fundamental Research, Homi Bhabha Road, Mumbai 400005, India}
\author{Shamik Gupta}
\email{shamik.gupta@theory.tifr.res.in}
\affiliation{Department of Theoretical Physics, Tata Institute of Fundamental Research, Homi Bhabha Road, Mumbai 400005, India}
\begin{abstract}
Stabilizing thermodynamically unstable phases in many-body systems, such as suppressing pathological neuronal synchronization in Parkinson’s disease or maintaining magnetic order across broad temperature ranges, remains a persistent challenge. In traditional approaches, such phases are stabilized through intervening in the dynamics of all system constituents or introducing additional interactions. Here, we offer a hitherto-unexplored alternative, namely, subsystem resetting, whereby intervention in the dynamics of only a part of the system, and that too only occasionally in time, is implemented through resetting its state to a reset configuration. Just playing with a few parameters, e.g., the nature of the reset configuration and the size of the reset subsystem, one achieves a remarkable and robust control over the phase diagram of the bare dynamics. We demonstrate that these universal effects span a wide variety of scenarios, including equilibrium and non-equilibrium, mean-field and non-mean-field dynamics, with and without quenched disorder. Despite the challenges posed by memory effects, we obtain explicit analytical predictions, validated by simulations. 
\end{abstract}
\maketitle
\def\thefootnote{$\mathsection$}\footnotetext{These authors contributed equally to this work.}\def\thefootnote{\arabic{footnote}}

As is well known, Kapitza’s seminal study of a pendulum with a rapidly oscillating suspension point demonstrated that dynamical perturbations can stabilize an otherwise unstable equilibrium. This allows the pendulum to undergo small oscillations around its inverted position, seemingly defying gravity~\cite{Kapitza1951}. This work serves as a cornerstone in dynamical system studies~\cite{Landau}, inspiring a wide range of applications aimed at stabilizing unstable states through dynamical intervention, spanning fields such as quantum physics~\cite{PhysRevB.100.104306,PhysRevLett.111.090403,PhysRevLett.104.240402}, atomic physics~\cite{Bukov04032015,bagnato1994dynamical,PhysRevLett.108.020603}, plasma physics~\cite{10.1063}, and biophysics~\cite{weinberg2014high, tass2007phase}. 
Remarkably, analogous concepts involving controlled dynamical interventions have been successfully employed in a clinical setting, whereby targeted stimulation has been proposed to drive neurons in Parkinson’s patients away from a state of pathological synchronization toward a healthy desynchronized state~\cite{tass2003model,tass2007phase,tass2011device}.

Inspired by both classic and contemporary ideas, we propose an alternative approach to stabilizing an unstable phase of a many-body interacting system by introducing occasional dynamical interventions in a small part of the system. The proposed protocol involves subsystem resetting~\cite{PhysRevE.109.064137, zhao2024subsystem, singh2024competing, bressloff2024kuramoto}, wherein the dynamics of the system is intervened repeatedly at random times at which a subpart of the system is reset to the desired state, while the rest evolves undisturbed. Between successive resets, the system follows its bare dynamics. Resetting, an active area of research~\cite{PhysRevLett.106.160601, Evans_2020,gupta2022stochastic,nagar2023stochastic, pal2022inspection, evans2024stochasticresettinglargedeviations}, has been studied across domains: classical~\cite{evans2013optimal,PhysRevLett.113.220602,PhysRevLett.112.220601, PhysRevE.92.062148,christou2015diffusion, pal2016diffusion,PhysRevE.96.022130,majumdar2018spectral,PhysRevE.99.012121,boyer2019anderson,PhysRevE.100.032136,PhysRevE.100.042104,chechkin,karthika2020totally,PhysRevResearch.2.033182,ciliberto,PhysRevResearch.2.043390,PhysRevE.101.062147,rosemary,gupta2021resetting,de2021optimization,evans2022exactly,rksingh,10.1063/5.0090861,PhysRevE.105.044134,PhysRevE.106.044127,10.1063/5.0196626,PhysRevE.110.014116,PhysRevResearch.6.033189, aron2024controlspatiotemporalchaosstochastic,PhysRevE.106.034137,PhysRevE.105.L012106,PhysRevResearch.4.013161,10.1063/5.0255601}, quantum~\cite{PhysRevB.98.104309,PhysRevB.104.L180302,fazio,das2022quantum,PhysRevE.108.064125,PhysRevLett.130.050802,PhysRevA.108.062210}, chemical~\cite{PhysRevE.92.060101,PhysRevResearch.3.013273},  biological~\cite{ramoso2020stochastic}, financial~\cite{stojkoski2022income,jolakoski2023first}.  Let us define reset subsystem as the part undergoing resetting and the rest as the non-reset subsystem. We ask: How can we change the amount of order, the nature of transition and transition points in the non-reset subsystem by tuning (i) the size of the reset subsystem, (ii) how often reset happens, and (iii) the nature of the reset configuration?

We address our queries in diverse dynamical setups, namely, two Hamiltonian systems involving $N$ discrete classical spins $s_i;~i=1,2,\ldots,N$: the Blume-Emery-Griffiths (BEG) model~\cite{PhysRevA.4.1071,PhysRevLett.87.030601} (mean-field), involving globally-coupled spin-$1$'s, $s_i = 0,\pm1$, and the Kardar-Nagle (KN) model~\cite{PhysRevA.2.2124,PhysRevB.28.244,PhysRevLett.95.240604} (non-mean-field), with spin-$1/2$'s, $s_i = \pm 1$, on a one-dimensional periodic lattice. The third  setup is non-Hamiltonian: the noisy Kuramoto model of $N$ globally-coupled limit-cycle oscillators with phase $0 \le \theta_i < 2\pi;~i=1,2,\ldots,N$ and quenched-disordered frequencies~$\omega_i$~\cite{kuramoto1984chemical,sakaguchi1988cooperative,campa2020phase}; it also represents continuous classical $XY$ spins with mean-field interactions, driven out-of-equilibrium~\cite{gupta2018statistical}. The BEG and KN models relax to equilibrium, while the Kuramoto model attains a non-equilibrium steady state (NESS). Upon tuning the coupling parameter, they all exhibit both continuous and first-order transitions. With resetting, when they all attain an NESS, we study how subsystem resetting affects the bare-model phase transitions.

\begin{figure}
    \centering
\includegraphics[width=8.7cm]{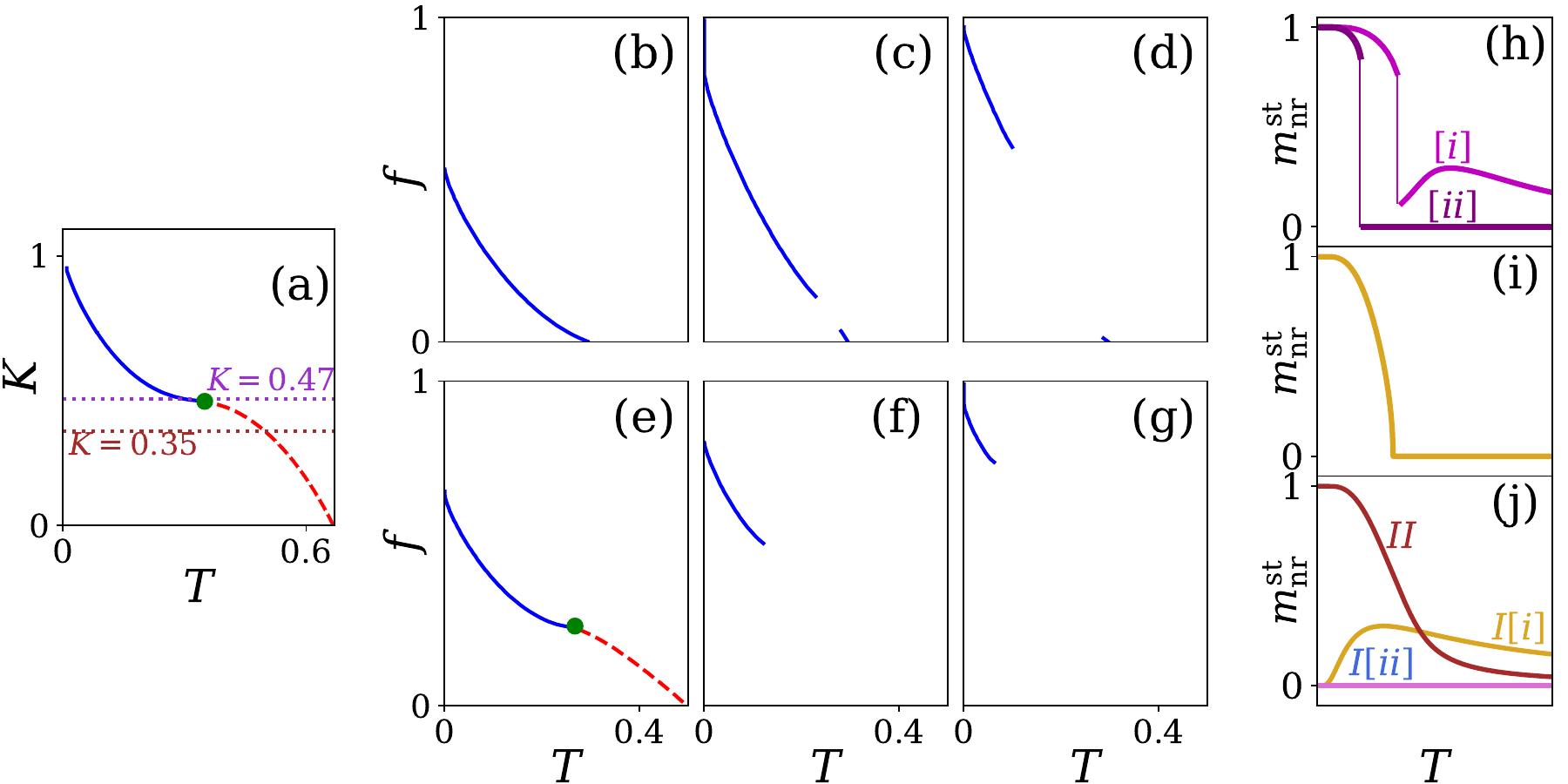}
    \caption{Phase diagram of BEG model~\eqref{eq:BEG Hamiltonian} in (a) $(K,T)$-plane without resetting, (b) -- (g) $(f,T)$-plane with subsystem resetting to configurations with varying order $m_0$ at rate $\lambda \to \infty$ and $K = 0.47$ (b) -- (d), $K = 0.35$ (e) -- (g), using~\eqref{eq:non reset magnetisation}; $m_0=0,0.2,0.3$ (b) -- (d) and $m_0=0,0.36,0.4$ (e) -- (g). In this Letter, red-dashed and blue-solid lines indicate  continuous and first-order transitions, respectively. Schematic non-reset order parameter $m^\mathrm{st}_\mathrm{nr}$ vs. $T$ across (1) a first-order transition line in the phase diagrams: panel (h), with $[i]$ (respectively, $[ii]$) corresponding to $m_0 \neq 0$ (respectively, $m_0 = 0$);~(2) a continuous transition line in the phase diagrams: panel (i);~(3) a region without transition: panel (j), with $I$ and $II$ corresponding respectively to whether the region lies above and below a first-order-transition region: $I[i]$ (respectively, $I[ii]$) is for $m_0 \neq 0$ (respectively, $m_0 = 0$). 
}
\label{fig:1}
\end{figure}

In this Letter, we unveil that subsystem resetting has two dramatic and remarkable effects. Firstly, it reproduces the full phase diagram of the bare model without tuning its couplings, a striking result shown in Fig. \ref{fig:1}(e) for the BEG model and discussed later for the other models. Secondly, it systematically modifies the phase transitions of the bare model, exemplified in Figs. \ref{fig:1}(b)–\ref{fig:1}(g). These behaviors, observed across the studied diverse dynamical setups, demonstrate their ubiquity both in and out of equilibrium, encompassing mean-field and non-mean-field dynamics, as well as systems with and without disorder. The effectiveness of our protocol hinges on the reset subsystem having an extensive number of connections with the non-reset subsystem. Hence, our results apply to general long-range interacting systems~\cite{campa2014physics}. In short-range systems, the reset subsystem  induces only boundary effects in the non-reset subsystem that disappear in the limit of large systems.

From an analytical perspective, when compared to global resetting in which the entire system undergoes reset events, subsystem resetting poses a greater challenge as the non-reset subsystem retains memory of the entire time evolution. This inherent memory effect precludes direct application of renewal theory, a standard tool widely employed in resetting studies~\cite{Evans_2020}. Despite these challenges, we are able to derive both exact and approximate analytical results, validated through numerical simulations.

Subsystem resetting was previously studied~\cite{PhysRevE.109.064137} in the noiseless Kuramoto model for particular (e.g., Lorentzian and Gaussian) frequency distributions, through the application of the Ott-Antonsen ansatz~\cite{Ott_2008, Ott_2009}. This influential ansatz yields a tractable low-dimensional description of the dynamics of the order parameter in the limit $N \to \infty$. However, the ansatz is inherently restrictive: it applies only to a specific invariant manifold of initial conditions and ceases to hold in presence of noise. Furthermore, Ref.~\cite{PhysRevE.109.064137} addressed only resetting to the fully-synchronized state. In contrast, the present work significantly broadens the scope in three principal directions: (i) we develop a unified analytical framework that remains valid both with and without noise and accommodates arbitrary frequency distributions; (ii) we generalize resetting to encompass states with arbitrary degrees of synchrony, enabling rich phase control not considered in the earlier work; and (iii) we extend the analysis to a broad class of non-oscillator systems, thereby establishing the generality of subsystem-resetting phenomena.

\begin{figure}
\centering
\includegraphics[width=8.3cm]{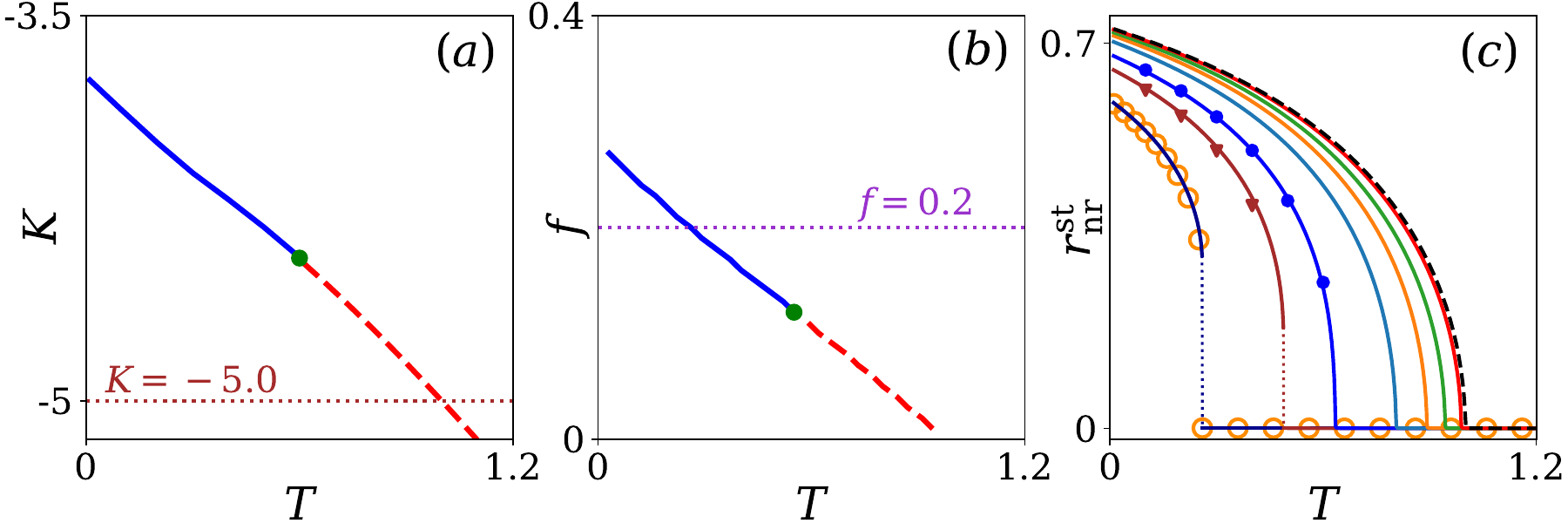}
\caption{Phase diagram of noisy Kuramoto model~\eqref{eq: Kuramoto Dynamics} with bimodal-Lorentzian frequency distribution $(\omega_0 = \sigma = 1)$ in (a) $(K,T)$-plane without resetting, (b) $(f,T)$-plane with subsystem resetting to a fully-disordered configuration at rate $\lambda \to \infty$ and  $K=-5.0$. For $(f,K)=(0.2,-5.0)$,~(c) shows order parameter versus $T$ changing from a continuous to a first-order transition as $\lambda$ is increased: $\lambda = 0.0, 0.1, 0.5, 1.0, 2.0, 5.0, 10.0, 5\times 10^2$, right to left. Lines: analytical results for finite $\lambda$~\eqref{eq: Finite Kuramoto SelfCon1}; filled points: simulation results with $N = 10^5$ oscillators; unfilled markers: $\lambda \to \infty$ limit of~\eqref{eq: Finite Kuramoto SelfCon1}.}
     \label{fig:4}
\end{figure}

We now turn to details. The BEG and KN Hamiltonians are
\begin{align}
    H_\textrm{BEG}&=K\sum\limits_{i=1}^N s_i^2-\frac{1}{2N}\sum_{i,j=1}^N s_is_j;~K>0, \label{eq:BEG Hamiltonian}\\
    H_\textrm{KN}&=\frac{K}{2}\sum_{i=1}^N (s_i s_{i+1}-1)-\frac{1}{2N}\sum_{i,j=1}^N s_is_j;~K>0.\label{eq:KN Hamiltonian}
\end{align}
In canonical equilibrium at temperature $T=1/\beta$ (Boltzmann constant $k_B=1$), Glauber dynamics~\cite{10.1063/1.1703954} models their time evolution. By contrast, the Kuramoto-model phases evolve as 
 \begin{eqnarray}
     \frac{d\theta_i}{dt} = \omega_i - \frac{K}{N} \sum_{j=1}^N \sin\left( \theta_j-\theta_i\right) + \zeta_i(t);~K<0, \label{eq: Kuramoto Dynamics}
 \end{eqnarray}
with Gaussian, white noise $\zeta_i(t)$ satisfying $\langle \zeta_i (t) \rangle = 0$, $\langle \zeta_i (t) \zeta_j (t') \rangle= \sqrt{2T}\, \delta_{ij}\delta(t-t')$, and $\omega_i$'s following a bimodal-Lorentzian distribution $g(\omega) = (\sigma/2\pi)\{1/[(\omega-\omega_0)^2+\sigma^2]+1/[(\omega+\omega_0)^2+\sigma^2]\}$.
 All the three models exhibit order-disorder phase transitions: (i) BEG and KN models in equilibrium, from a low-$T$ ferromagnetic ($m\ne0$) to a high-$T$ paramagnetic ($m=0$) phase in magnetization order parameter $m \equiv  (1/N)\sum_{i=1}^N s_i$, and (ii) Kuramoto model in NESS, from  a low-$T$ synchronized $(r \neq 0)$ to a high-$T$ incoherent $(r=0)$ phase in synchronization order parameter $re^{i\psi} \equiv  (1/N) \sum_{j=1}^N e^{i \theta_j}$. The transition changes from continuous to first-order on increasing $K$ [Figs.~\ref{fig:1}(a),~\ref{fig:4}(a) ,~\ref{fig:5}(a) (Appendix A)]. Note that $m=r=0$ represents fully disordered phase in respective models, and all the models have long-range interaction (this is why the KN model shows phase transitions despite being in $1d$ where transitions are precluded with sole short-range interactions).   

We define the subsystem resetting protocol~\cite{PhysRevE.109.064137} for all the models: among $N$ constituents (spins/oscillators), $n<N$ of them (labeled $i = 1,2,\ldots,n$), chosen uniformly and independently, form the reset (r) subsystem (with size $f \equiv n/N$), and only these undergo resetting. The remaining constituents form the non-reset (nr) subsystem with size $\bar{f}=1-f$. Along with global order parameters $(m,r)$, we define order parameters for individual subsystems: $m_\mathrm{r} \equiv  (1/n)\sum_{i=1}^n s_i$,  $r_\mathrm{r}e^{i \psi_\mathrm{r}} \equiv (1/n) \sum_{j=1}^n e^{i \theta_j} $, $m_\mathrm{nr} \equiv [1/(N-n)]\sum_{i=n+1}^{N} s_i$,  and $r_\mathrm{nr}e^{i \psi_\mathrm{nr}} \equiv [1/(N-n)] \sum_{j=n+1}^{N} e^{i \theta_j} $. Let the system be initiated at fully-ordered configuration ($m=r=1$). The dynamics with resetting involves bare evolution (Glauber dynamics for Hamiltonian~\eqref{eq:BEG Hamiltonian} and~\eqref{eq:KN Hamiltonian} or dynamics~\eqref{eq: Kuramoto Dynamics}) repeatedly interrupted at exponentially-distributed random time intervals (with rate $\lambda>0$), whereby the reset subsystem is reset to a given configuration called the reset-configuration (its order parameter resetting to corresponding value of the latter: $m_\mathrm{r}$ to $m_0$, $r_\mathrm{r}$ to $r_0$ and $\psi_{r}$ to $\psi_0$), while the non-reset subsystem is left unaltered ($m_\mathrm{nr},r_\mathrm{nr},\psi_\mathrm{nr}$ unchanged during resets). The reset instants (not the reset subsystem) vary across dynamical realizations. The limit $\lambda \to 0$ recovers bare dynamics, while $\lambda \to \infty$ maximizes resetting effects for given $f$, system parameters $(K, T$ or $K, \omega_0, \sigma, T)$, reset values $(m_0$ or $(r_0, \psi_0))$.

We present detailed results for the BEG model and representative ones for the KN and Kuramoto models; despite their dissimilarities, they yield qualitatively-similar results (see Appendices). For both $\lambda \to \infty$ and finite-$\lambda$, changing $f$ or $m_0 (r_0)$ at fixed $\lambda$ or changing $\lambda$ for fixed $f,m_0 (r_0)$ allows to manipulate the phases. Moreover, $\lambda\to \infty$-results are achieved with finite but not-too-large $\lambda$ ($=10.0$ for BEG and KN models). 

In the BEG model, as $\lambda \to \infty$, the reset subsystem is frozen at the reset configuration,  $m_\mathrm{r}=m_0$ (equivalent of the Zeno limit in quantum dynamics~\cite{misra1977zeno,kulkarni2023first}); Eq.~\eqref{eq: BEG eff Ham Non Reset} then gives the effective Hamiltonian (up to an additive constant) of non-reset subsystem with configuration $\mathcal{C}\equiv \{s_{n+1},\ldots, s_N\}$: 
\begin{align}
    \hspace{-0.25cm}H^{\lambda \to \infty}_\mathrm{BEG} = K\sum_{s_i \in \mathcal{C}} s_i^2 -  f m_0\sum_{s_i \in \mathcal{C}} s_i - \frac{1}{2N} \sum_{s_i,s_j \in \mathcal{C}} s_is_j.
    \label{eq: BEG eff Ham Non Reset}
\end{align}
The Hubbard-Stratonovich (HS) transformation yields canonical partition function $Z_{\mathrm{BEG}}=\sum_{\mathcal{C}} \exp(-\beta H_{\mathrm{BEG}}^{\lambda \to \infty})=\sqrt{\beta  N/(2 \pi)}\bar{f}\int_{-\infty}^{\infty} d m_{\mathrm{nr}}\exp(-N \beta \tilde{F}(\beta,m_{\mathrm{nr}},f))$, with $\tilde{F}$ $\equiv (\bar{f}^2/2)m^2_\mathrm{nr}-(\bar{f}/\beta) \ln \left\{1+ 2 e^{-\beta K}\cosh[\beta  (fm_0 +\bar{f}m_\mathrm{nr})]\right\}$. As $N\to \infty$ (thermodynamic limit), evaluating the integral by saddle point yields free-energy/spin as $\tilde{F}(\beta,m^\mathrm{st}_{\mathrm{nr}},f)$, with steady-state magnetization $m^\mathrm{st}_{\mathrm{nr}}$ (the particular $m_{\mathrm{nr}}$ that minimizes $\tilde{F}(\beta,m_{\mathrm{nr}},f)$) satisfying~(Supplemental Material~\cite{supplement})
\begin{align}
     \hspace{-0.25cm}m^\mathrm{st}_\mathrm{nr} = \frac{2 \sinh\left[\beta  (f m_0 + \bar{f}m^\mathrm{st}_\mathrm{nr})\right]}{e^{\beta K}+2 \cosh\left[\beta  ( f m_0 + \bar{f}m^\mathrm{st}_\mathrm{nr})\right]}; |m_0|\le 1;\label{eq:non reset magnetisation}
\end{align}
$\mathbb{Z}_2$ symmetry of $H_\mathrm{BEG}$ allows restricting to $0\leq m_0 \leq 1$. We now present explicit results for $\lambda \to \infty$, based on~\eqref{eq:non reset magnetisation}

\begin{figure}
    \centering
\includegraphics[width=8.2cm]{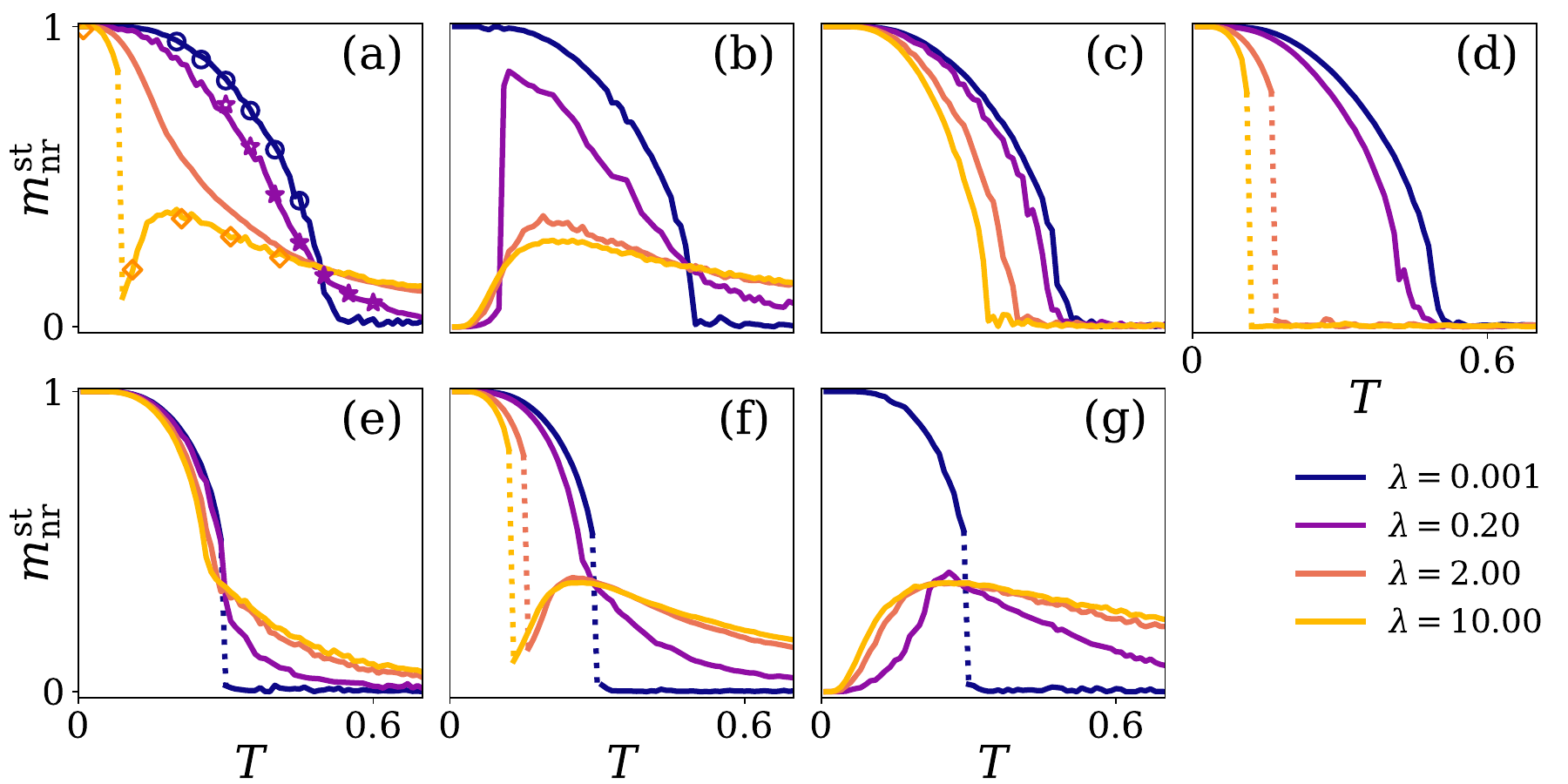}
    \caption{For the BEG model~\eqref{eq:BEG Hamiltonian} with $K=0.35$~(a) -- (d) and $K=0.47$~(e) -- (g), the figure shows the behavior of $m^\mathrm{st}_\mathrm{nr}$ versus $T$, changing with increase of $\lambda$ from that of bare model ($\lambda \to 0$) to that in the limit $\lambda \to \infty$: $\lambda = 0.001, 0.20, 2.00, 10.00$, right to left in each panel. The data are obtained from simulations for $N=8\times 10^3$ spins; results for $\lambda=10.0$ coincide with $\lambda\to\infty$ analytical results. The panels correspond to different values of $f$ and $m_0$: (a),~(b): $f = 0.6, 0.9$ and $m_0 = 0.2$; (c),~(d): $f = 0.2, 0.4$ and $m_0 = 0$; (e),~(f),~(g): $f =0.1, 0.4, 0.9$ and $m_0 = 0.36$. Unfilled markers in (a) are theoretical estimates using~\eqref{eq: estimate BEG finite}.}
    \label{fig:2}
\end{figure}

\textit{(A) $K$-values $(0 \leq K <0.4621)$ with continuous transition in bare-BEG model} (Figs.~\ref{fig:1}(e)-\ref{fig:1}(g)): $m_0=0$ at fixed $K$ generates in $(f,T)$-plane qualitatively the entire phase-diagram of the bare-BEG model, Fig.~\ref{fig:1}(e). With respect to (e), as $m_0$ is increased, the continuous transition gets replaced by a crossover, whose region expands and eventually spans the whole $(f,T)$-plane, subsuming the first-order-transition region and the crossover region above it. \textit{(B) $K$-values $(0.4621 \leq K \leq 0.95)$ with first-order transition in bare-BEG model} (Figs.~\ref{fig:1}(b) -\ref{fig:1}(d)): In contrast to above, here subsystem resetting retains the first-order transition of the bare model or converts it into a crossover: For $m_0=0$ as also for small $m_0$, at fixed $K$, the first-order-transition region of the bare-BEG splits into a crossover and a first-order-transition region in $(f,T)$-plane. With increasing $m_0$, the latter splits further into two first-order transition regions with a crossover in between. At higher $m_0$, the latter expands, eventually spanning the $(f,T)$-plane while retaining the lower first-order transition region. Figures~\ref{fig:1}(h) -\ref{fig:1} (j) shows schematic $m_\mathrm{nr}^{\mathrm{st}}$ versus $T$. Without resetting, any subsystem has the bare-model phase diagram in the thermodynamic limit. Interestingly, tuning the reset-subsystem size alters both the nature and point of transitions in the $(f,T)$-phase diagram of the non-reset part, with systematic changes as $m_0$ increases.

\begin{figure}
    \centering
    \includegraphics[width=8.7 cm]{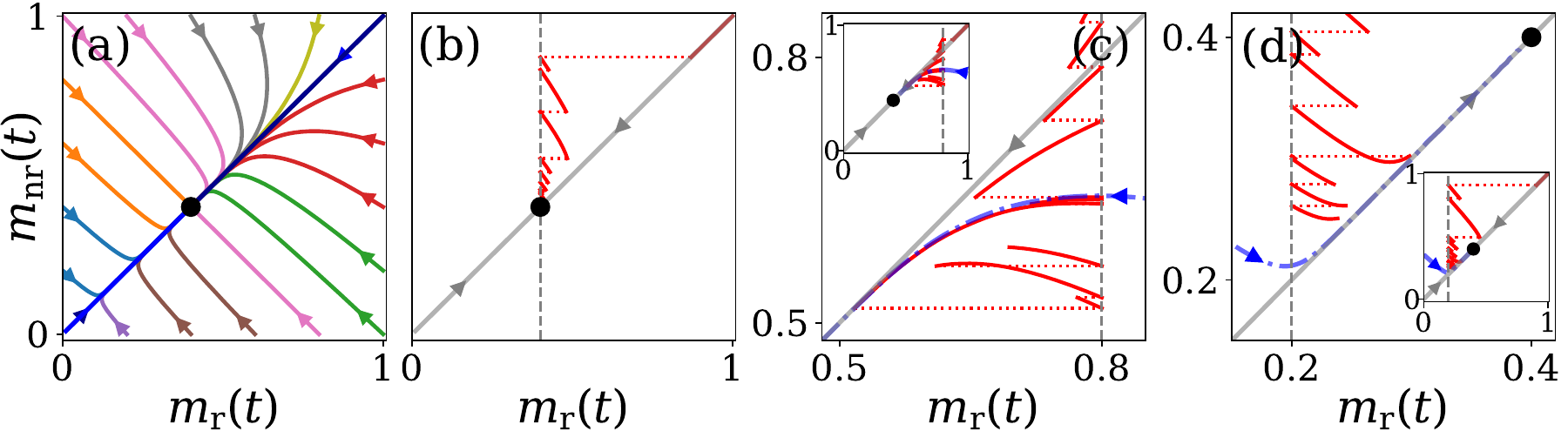}
   \caption{(a) Flow diagram of bare BEG model in $(m_\mathrm{r},m_\mathrm{nr})$-plane for $K=0.35,~f=0.5,~T=0.455$. The black circle denotes the steady state, while lines with arrows show convergence to it from different initial conditions. The following three plots depict the scenario in presence of resetting, with $m_0 = 0.4, 0.8, 0.2$ (b) -- (d). The vertical dashed line stands for $m_\mathrm{r}=m_0$, while the dot-dashed line in (c) and (d) denotes the inflection line (in each case, the main plot is a zoom-in onto the inset). In each plot, the red line shows a typical dynamical trajectory starting from $(m_\mathrm{r},m_\mathrm{nr})=(1,1)$, with reset events denoted by dotted horizontal lines.}
     \label{fig:3}
\end{figure}

For finite $\lambda$, when reset and non-reset subsystems have different dynamics with no time-scale separation, Fig.~\ref{fig:2} shows simulation results on $m^\mathrm{st}_{\mathrm{nr}}$ versus $T$ interpolating between the bare model ($\lambda \to 0$) and $\lambda\to \infty$ results. Relevant observations are: (i) On increasing $\lambda$, provided $m_0 \neq 0$, transitions of the bare model convert into crossover and then into the transitions/crossover for $\lambda\to \infty$-limit (for $m_0=0$, transitions sustain for all $\lambda$,~Figs.~\ref{fig:2}(c) -\ref{fig:2}(d)). (ii) $m^\mathrm{st}_\mathrm{nr}$ - $T$ plots for fixed $m_0$ and different $\lambda$ intersect at $(m^\mathrm{st}_\mathrm{nr},T)=(m_0,\overline{T}$). (iii) For $T<\overline{T}$ (respectively, $T>\overline{T}$), $m^\mathrm{st}_\mathrm{nr}$ in presence of resetting is smaller (respectively, larger) than in the bare model. To explain these features, we analyze the bare-BEG flow-diagram in $(m_\mathrm{r}, m_\mathrm{nr})$-plane. Glauber dynamics yields~\cite{supplement} $d m_\mathrm{x}/d t   = - m_\mathrm{x} +\{ 2 \sinh{[\beta J (f m_\mathrm{r} + \bar{f}m_\mathrm{nr})]} \} /\{2 \cosh{[\beta J (f m_\mathrm{r} + \bar{f}m_\mathrm{nr})]} + e^{\beta K} \}$; $\mathrm{x}=\mathrm{r},\mathrm{nr}$, generating for the initial condition $m^{(0)}_\mathrm{x}$ the flow 
in time as $m_\mathrm{x}(t| m^{(0)}_\mathrm{r},m^{(0)}_\mathrm{nr})$.  Every spin having identical dynamics implies that any stable fixed point(s), denoting steady-state, lies on $m_\mathrm{r}=m_\mathrm{nr}$-line. Out of the two low-$T$ stable fixed points, at $(0,0)$ and close to $(1,1)$, the former vanishes on increasing $T$, while the latter shifts down the $m_\mathrm{r}=m_\mathrm{nr}$-line. Considering a temperature with one stable point $(m^T_\mathrm{bare},m^T_\mathrm{bare})$, Fig.~\ref{fig:3}(a), all flow lines (the dynamics being first order, flow lines are non-intersecting) first approach the $m_\mathrm{r}=m_\mathrm{nr}$-line before converging to the stable point. The ``inflection" flow-line, defined as $dm_\mathrm{nr}/d m_\mathrm{r}|_{m_\mathrm{r}=m_0} = 0$, depends on $m_0$. If the dynamics is initiated with $m_\mathrm{r}=m_0$ and $m_\mathrm{nr}$ arbitrary, then, near the inflection line, $m_\mathrm{nr}$ converges towards $m^T_\mathrm{bare}$ monotonically on one side (namely, above the line in Fig.~\ref{fig:3}(c) and below in ~\ref{fig:3}(d)) and non-monotonically on the other side (respectively, below in Fig.~\ref{fig:3}(c) and above in ~\ref{fig:3}(d)).

We now discuss resetting effects at finite $\lambda$ for (I) $m_0= m^T_\mathrm{bare}$, (II) $m_0 > m^T_\mathrm{bare}$, and (III) $m_0 < m^T_\mathrm{bare}$. Figures~\ref{fig:3}(b) -\ref{fig:3} (d) shows a typical dynamical trajectory initiated at $(1,1)$ (red line), and that resetting repeatedly shifts the dynamics from one flow line to another. For (I), Fig.~\ref{fig:3}(b), the trajectory remains confined between the lines $m_\mathrm{r}=m_\mathrm{nr}$ and $m_\mathrm{r} = m_0$, converging to the bare-model stable point. Thus, $m^\mathrm{st}_\mathrm{nr} = m^T_\mathrm{bare}=m_0$ for any $\lambda$. This argument holds provided for any given $m_0$, one finds a $T$ such that $m^T_\mathrm{bare}=m_0$. This is true, except when the bare model exhibits a first-order transition with $m^T_\mathrm{bare}$ exhibiting a jump (a gap) in its value and $m_0$ has a value within the gap; yet, even then, our simulations show that observation (ii) holds for the BEG model. For (II), Fig.~\ref{fig:3}(c), the trajectory when confined above the inflection line results in $m_\mathrm{nr}$ decreasing monotonically, due to the feature of the inflection line mentioned above. Once the trajectory crosses the inflection line, the non-monotonic flow below the line, together with resetting events, confines the trajectory between the inflection, the $m_\mathrm{r}=m_\mathrm{nr}$ and the $m_\mathrm{r}=m_0$ line. This results in $m^\mathrm{st}_\mathrm{nr} > m^T_\mathrm{bare}$. For  (III), a similar reasoning implies $m^\mathrm{st}_\mathrm{nr} < m^T_\mathrm{bare}$. To estimate $m^\mathrm{st}_\mathrm{nr}$, considering resetting at regular interval $\tau$ and that evolution between two resets follows the bare dynamics, the steady state in presence of resetting follows the fixed-point equation $m_\mathrm{nr}\left(\left.\tau \right| m_0, y\right) = y$. Resetting at random intervals, with average $\langle \tau \rangle=1/\lambda$, estimates the steady-state magnetization $y$ to be satisfying
\begin{eqnarray}
    m_\mathrm{nr}\left(\left.1/\lambda \right| m_0, y\right) = y, \label{eq: estimate BEG finite}
\end{eqnarray} 
whose solution matches well with simulations,  Fig.~\ref{fig:2}(a). Herewith, we have explained the relevant features of Fig.~\ref{fig:2}.

For the KN model, the effective Hamiltonian of the non-reset subsystem with configuration $\mathcal{C}$ as $\lambda \to \infty$ is
\begin{eqnarray}
    H_{\mathrm{KN}}^{\lambda \to \infty} 
   &&= \frac{K}{2}\sum\limits_{s_i \in \mathcal{C}}(s_i s_{i+1}-1)-fm_0 \sum_{s_i \in \mathcal{C}} s_i\nonumber\\
    &&\hspace{-1 cm}-\frac{1}{2N} \sum_{s_i, s_j \in \mathcal{C}}s_is_j -\frac{K}{2} (s_N-1)(s_{n+1}-1).
    \label{eq:HKN-lambda-infinity}
\end{eqnarray}
Being non-mean-field, evaluating the canonical partition function  requires different treatments than the BEG model. The  HS transformation yields the partition function as $Z_{\mathrm{KN}}=\sqrt{\beta  N/(2 \pi)} \bar{f}\int_{-\infty}^{\infty} d m_{\mathrm{nr}}~\exp(-N \beta \tilde{F}(\beta,m_{\mathrm{nr}},f))$; $ \tilde{F}(\beta,m_{\mathrm{nr}},f)\equiv (\bar{f}^2/2)m_{\mathrm{nr}}^2 + K\bar{f}  - F_0 (\beta, m_{\mathrm{nr}} ,f)$, and $F_0$ the free-energy of the nearest-neighbor Ising model of $N-n$ spins in an external field of strength $M \equiv f m_0+\bar{f} m_{\mathrm{nr}} $, and with two additional boundary terms. Evaluating the latter by transfer matrix finally yields in the thermodynamic limit~\cite{supplement} 
\begin{equation}
    m^\mathrm{st}_\mathrm{nr}=  \frac{\sinh[\beta \left( f m_0 + \bar{f} m^\mathrm{st}_\mathrm{nr} \right)]}{\sqrt{ \cosh^2[\beta \left( f m_0 + \bar{f} m^\mathrm{st}_\mathrm{nr}\right)] + 2 e^{\beta K} \sinh{\beta K}}} \label{eq: KN Inf Reset Self Con}.
\end{equation}
Results based on~\eqref{eq: KN Inf Reset Self Con} show similar features as in the BEG model as $\lambda \to \infty$. For finite-$\lambda$, the corresponding observation (iii)  for the BEG model also holds, while observations (i) and (ii) hold in the case of continuous transitions in bare KN, with differences when the transition is first order for $m_0 \neq 0$ (Appendixes A and B in Fig.~\ref{fig:6}). Remarkably, for $m_0 = 0$, the model behaves similarly to the BEG for both finite and infinite $\lambda$.

In Kuramoto model, due to its non-equilibrium nature, the finite-$\lambda$ analysis is most nontrivial compared to the BEG and KN models, which we report here and from which one can recover the $\lambda \to \infty$-results. The latter may also be derived by invoking a mapping to an effective model, like the BEG and KN models. For finite-$\lambda$ and $N \to \infty$, the dynamics may be characterized by the joint probability density $P(\theta_\mathrm{r}, \theta_\mathrm{nr}, \omega_\mathrm{r}, \omega_\mathrm{nr}, t)$, with $(\theta_\mathrm{r}, \omega_{r})$ and $(\theta_\mathrm{nr}, \omega_{nr})$ being respectively the angle and frequency of an oscillator from the reset and the non-reset subsystem; its time evolution follows
\begin{eqnarray}
    \nonumber \frac{\partial P}{\partial t} =&& T \left[\frac{\partial^2 P}{\partial \theta^2_\mathrm{r}} +  \frac{\partial^2 P}{\partial \theta^2_\mathrm{nr}}\right] -\left[ \frac{\partial \left(P h_\mathrm{r} \right)}{\partial \theta_\mathrm{r}}+ \frac{\partial \left(P h_\mathrm{nr} \right)}{\partial \theta_\mathrm{nr}} \right] -\lambda P\\
    &&+ \lambda \left[ \alpha \delta(\theta_\mathrm{r}) + (1-\alpha) \delta(\theta_\mathrm{r}-\pi)\right]\nonumber\\
    && \times\int_{-\infty}^{+\infty} d \omega'_\mathrm{r} g(\omega'_\mathrm{r})\int_0^{2\pi} d \theta'_\mathrm{r} P(\theta'_\mathrm{r},\theta_\mathrm{nr},\omega'_\mathrm{r},\omega_\mathrm{nr},t); \label{eq: FP Kuramoto Finite Resetting} 
\end{eqnarray}
 $h_\mathrm{x} = \omega_\mathrm{x} -Kf \int d\theta'_\mathrm{r}d\omega'_\mathrm{r} g(\omega_\mathrm{r}) P(\theta'_\mathrm{r},\omega'_\mathrm{r} ,t| \theta_\mathrm{x} ,\omega_\mathrm{x})$ $\sin(\theta'_\mathrm{r}-\theta_\mathrm{x})-K\bar{f} \int d\theta'_\mathrm{nr}d\omega'_\mathrm{nr} P(\theta'_\mathrm{nr},\omega'_\mathrm{nr} ,t|  \theta_\mathrm{x},\omega_\mathrm{x}) \sin(\theta'_\mathrm{nr}-\theta_\mathrm{x})$, 
 with $\mathrm{x} = \mathrm{r}, \mathrm{nr}$ and $|2\alpha -1| = r_0$. The conditional probabilities involve the joint distribution $\mathbb{P}(\theta_\mathrm{r},\theta'_\mathrm{r}, \theta_\mathrm{nr}, \theta'_\mathrm{nr},\omega_\mathrm{r},\omega'_\mathrm{r},\omega_\mathrm{nr},\omega'_\mathrm{nr},t)$. The first two bracketed terms on the right of~\eqref{eq: FP Kuramoto Finite Resetting} are respectively the usual diffusion term due to Gaussian noise and the drift term due to inter-oscillator interactions, while the last two terms account for probability loss and gain due to resetting at rate $\lambda$. To implement resetting to $r_0$, an $\alpha$ fraction of the reset-subsystem oscillators are set to the phase-value zero and the remaining $(1-\alpha)$ to the value $\pi$, giving $r_0 = |2\alpha-1|$.  In the steady state, assuming $\mathbb{P}_\mathrm{st} \approx P_\mathrm{st}(\theta_\mathrm{r}, \theta_\mathrm{nr},\omega_\mathrm{r},\omega_\mathrm{nr}) P_\mathrm{st}(\theta'_\mathrm{r},  \theta'_\mathrm{nr},\omega'_\mathrm{r},\omega'_\mathrm{nr})$, we get~(Appendix C):
\begin{align}
 &r^\mathrm{st}_\mathrm{x}e^{i\psi^\mathrm{st}_\mathrm{x}} = 2 \pi i  \sum_{\omega_q } \mathrm{Res} \left.\left[g(\omega)A_\mathrm{x} (\omega, z^\mathrm{st}_\mathrm{r}, z^\mathrm{st}_\mathrm{nr}) \right]\right|_{\omega=\omega_q}, \label{eq: Finite Kuramoto SelfCon1}
\end{align}
with $A_\mathrm{r} \equiv \Gamma^{*}_1 + 4 \pi^2\Delta^{*}_1, A_\mathrm{nr} \equiv \Lambda_1^{*}, z^\mathrm{st}_\mathrm{x} \equiv r^\mathrm{st}_\mathrm{x}e^{i\psi^\mathrm{st}_\mathrm{x}} $, and $\omega_q$ being the poles of $g(\omega)$ in the lower-half of the complex-$\omega$ plane. The above expressions hold for any $g(\omega)$, and may be evaluated for our choice of bimodal-Lorentzian $g(\omega)$; the $\lambda \to \infty$-limit is discussed in~\cite{supplement}. These analytical results along with numerical verification are presented in Figs.~\ref{fig:4}(b), \ref{fig:4}(c) and in Appendixes D in Fig.~\ref{fig:7} and E in \ref{fig:8}. As $\lambda \to \infty$, we find similar behavior as that of BEG and KN models (Appendix D). For finite-$\lambda$, the corresponding observation (iii) made for the BEG model remains valid here; observations (i) and (ii) do not hold for either continuous or first-order transitions in the Kuramoto model for $r_0 \neq 0$ (Appendix E). For both finite and infinite $\lambda$, one has for $r_0 = 0$ a behavior similar to the BEG and KN models with $m_0=0$.

In summary, this work demonstrates that subsystem resetting serves as a versatile protocol for steering many-body systems toward desired states by simply adjusting the size of the reset subsystem, selecting appropriate reset configurations, and deciding on how often to reset. We solved exactly for effects of subsystem resetting in a variety of classical many-body systems. Notably, our findings reveal that resetting can replicate the complete phase diagram of the bare model without requiring fine-tuning of its couplings, while also enabling systematic manipulation of phase-transition points. Our work advances a new paradigm for manipulating collective dynamics, providing effective tools for design and regulation of complex systems. In complex systems such as the brain, interactions are typically heterogeneous and deviate from being mean-field. In this setting, resetting the influential driver nodes may offer an efficient control strategy even with small values of $f$. Indeed, in the case of heterogeneous networks~\cite{RODRIGUES20161,PhysRevLett.96.240602,PhysRevLett.97.100601,PhysRevE.79.011102}, we observe that resetting even a non-extensive number of nodes ($f \to 0$ as $N \to \infty$) can induce a measurable change in the synchronization behavior of the non-reset part of the network (see Appendix F). Our findings are amenable to experimental realization in long-range systems, particularly, in trapped-ion and cold-atom platforms~\cite{Katz_2023,vijayan2024cavity,defenu2023long} and non-reciprocal interacting systems~\cite{fruchart2021non}. 

\textit{Acknowledgments:} We acknowledge useful discussions with Julien Barr\'{e}, Shankar Ghosh, and Satya N. Majumdar, and especially thank Kedar Damle for critical comments on the manuscript. We gratefully acknowledge generous allocation of computational resources of the Department of Theoretical Physics, TIFR, assistance of Kapil Ghadiali and Ajay Salve, and financial support of Department of Atomic Energy, Government of India, under Project Identification No. RTI 4002.

\bibliography{main}
\bibliographystyle{unsrt}

\clearpage
\appendix
\section{End Matter}

\textit{Appendix A: $\lambda \to \infty$-results for KN model --}
 \begin{figure}[h!]
    \centering
\includegraphics[width=8 cm]{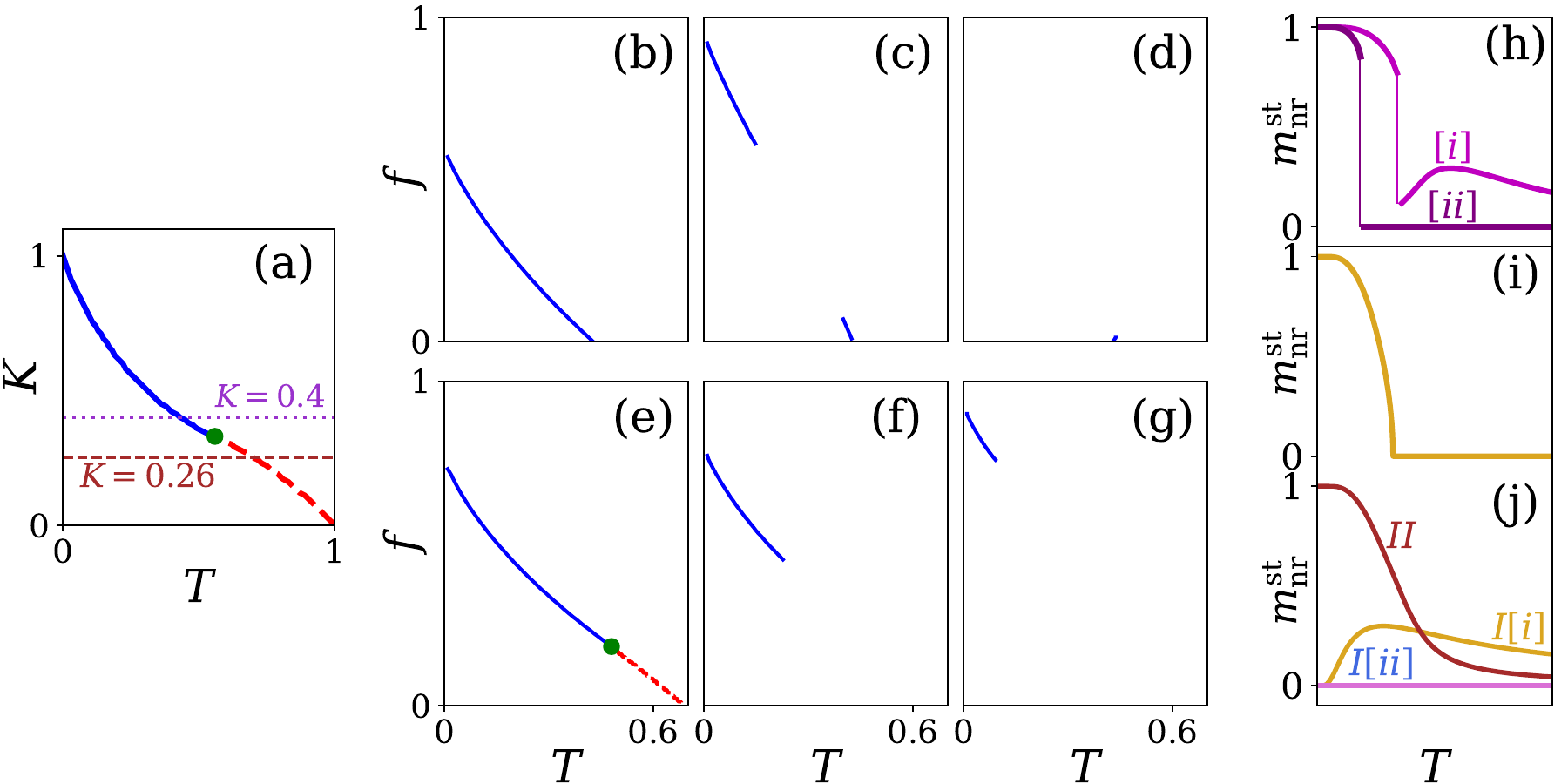}
    \caption{Phase diagram of KN model in (a) $(K,T)$-plane without resetting, (b) -- (g) $(f,T)$-plane with subsystem resetting to configurations with varying order $m_0$ at rate $\lambda \to \infty$ and $K = 0.4$ (b) -- (d), $K = 0.26$ (e) -- (g), using~(8) of main text; $m_0=0,0.37,1.0$ (b) -- (d) and $m_0=0,0.07, 0.2$ (e) -- (g). Schematic non-reset order parameter $m^\mathrm{st}_\mathrm{nr}$ vs. $T$ across (1) a first-order transition line in the phase diagrams: panel (h), with $[i]$ (respectively, $[ii]$) corresponding to $m_0 \neq 0$ (respectively, $m_0 = 0$);~(2) a continuous transition line in the phase diagrams: panel (i);~(3) a region without transition: panel (j), with $I$ and $II$ corresponding respectively to whether the region lies above or below a first-order-transition region: $I[i]$ (respectively, $I[ii]$) is for $m_0 \neq 0$ (respectively, $m_0 = 0$). 
}\label{fig:5}
\end{figure}

\textit{Appendix B: Finite-$\lambda$ results for KN model --}
\begin{figure}[h!]
    \centering
\includegraphics[width=8.0cm]{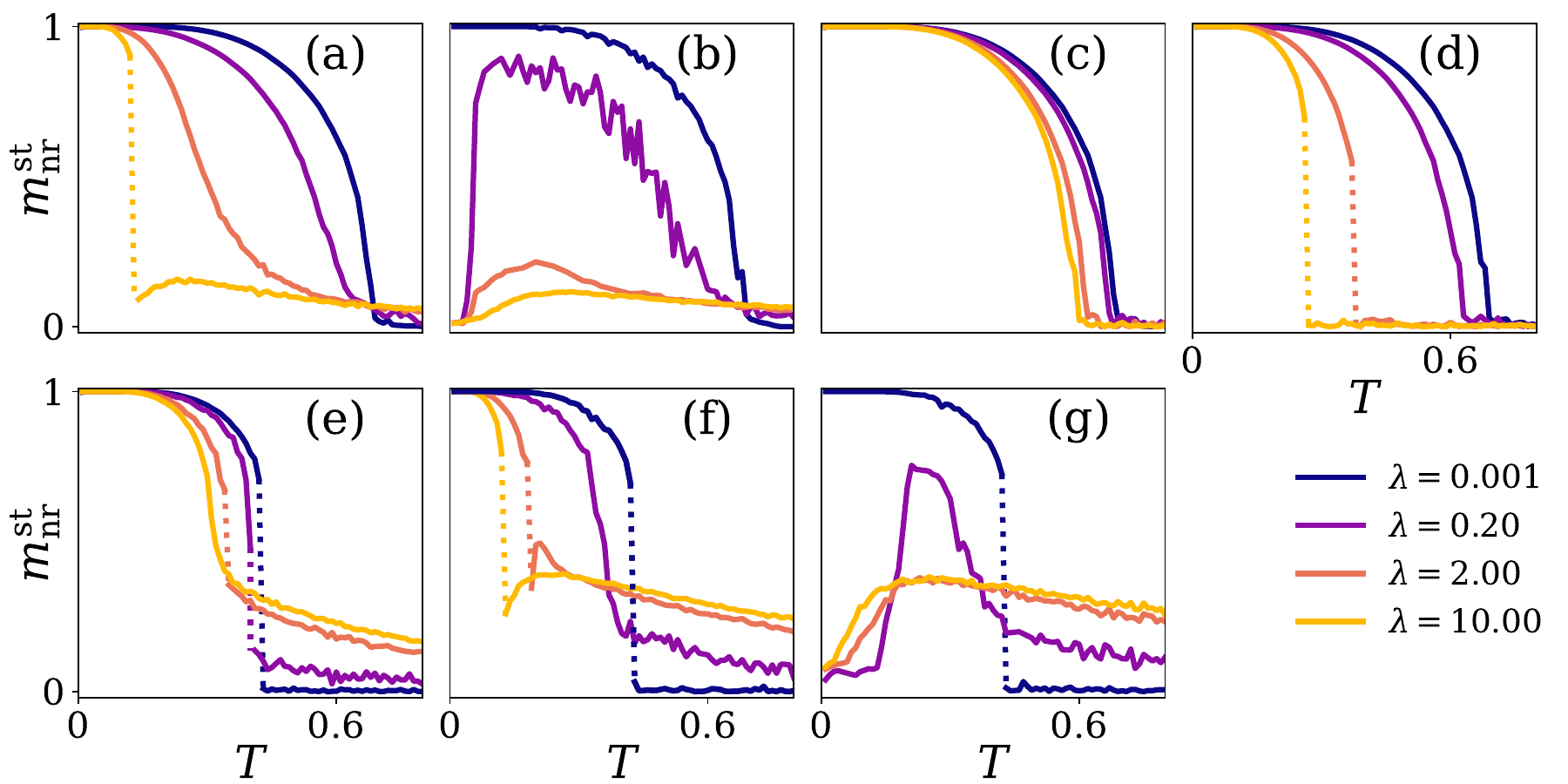}
    \caption{For the KN model~\eqref{eq:KN Hamiltonian} with $K=0.26$~(a) -- (d) and $K=0.40$~(e) -- (g), the figure shows the behavior of $m^\mathrm{st}_\mathrm{nr}$ versus $T$, with increase of $\lambda$, from that of bare model ($\lambda \to 0$) to that in the limit $\lambda \to \infty$: $\lambda = 0.001, 0.2, 2.0, 10.0$, right to left in each panel. The data are obtained from simulations for $N=8\times 10^3$ spins; results for $\lambda=10.0$ coincide with $\lambda\to\infty$ analytical results. The panels correspond to different values of $f$ and $m_0$: (a),~(b): $f = 0.60, 0.85$ and $m_0 = 0.07$; (c),~(d): $f = 0.10, 0.40$ and $m_0 = 0$; (e),~(f),~(g): $f =0.30, 0.70, 0.98$ and $m_0 = 0.37$. }\label{fig:6}
\end{figure}

\textit{Appendix C: Derivation of Eq.\;\eqref{eq: Finite Kuramoto SelfCon1}--}\label{sec:app0}

We start with the quantity $h_\mathrm{x}$ defined in the main text following Eq.~\eqref{eq: FP Kuramoto Finite Resetting}, which contains conditional probabilities of the form $P(\theta'_\mathrm{y},\omega'_\mathrm{y},t|\theta_\mathrm{x},\omega_\mathrm{x});~\mathrm{x},~\mathrm{y}=\mathrm{r},~\mathrm{nr}$. Now, these conditional probabilities contain the joint probability distribution $\mathbb{P}(\theta_\mathrm{r},\theta'_\mathrm{r}, \theta_\mathrm{nr}, \theta'_\mathrm{nr},\omega_\mathrm{r},\omega'_\mathrm{r},\omega_\mathrm{nr},\omega'_\mathrm{nr},t)$. In the steady state, assuming factorization $\mathbb{P}_\mathrm{st} (\theta_\mathrm{r},\theta'_\mathrm{r}, \theta_\mathrm{nr}, \theta'_\mathrm{nr},\omega_\mathrm{r},\omega'_\mathrm{r},\omega_\mathrm{nr},\omega'_\mathrm{nr}) \approx P_\mathrm{st}(\theta_\mathrm{r}, \theta_\mathrm{nr},\omega_\mathrm{r},\omega_\mathrm{nr}) P_\mathrm{st}(\theta'_\mathrm{r},  \theta'_\mathrm{nr},\omega'_\mathrm{r},\omega'_\mathrm{nr})$, it is straightforward to see that $P_\mathrm{st}(\theta'_\mathrm{y},\omega'_\mathrm{y}|\theta_\mathrm{x},\omega_\mathrm{x}) = P_\mathrm{st}(\theta'_\mathrm{y},\omega'_\mathrm{y})$.

Next, we exploit the periodicity $P(\theta_\mathrm{r}, \theta_\mathrm{nr}, \omega_\mathrm{r}, \omega_\mathrm{nr}, t) = P(\theta_\mathrm{r}+ 2\pi, \theta_\mathrm{nr}, \omega_\mathrm{r}, \omega_\mathrm{nr}, t) = P(\theta_\mathrm{r}, \theta_\mathrm{nr} + 2\pi, \omega_\mathrm{r}, \omega_\mathrm{nr}, t)$ , to expand $P_\mathrm{st}(\theta_\mathrm{r}, \theta_\mathrm{nr}, \omega_\mathrm{r}, \omega_\mathrm{nr})$ as a Fourier series:
\begin{align}
    P_\mathrm{st} = \sum_{l,m = -\infty}^{\infty} \mathcal{P}_{l,m}(\omega_\mathrm{r},\omega_\mathrm{nr}) e^{i l \theta_\mathrm{r}+i m\theta_\mathrm{nr}}. \label{eq: Fourier Expansion}
\end{align}
Now, $P_\mathrm{st}(\theta_\mathrm{r},\theta_\mathrm{nr},\omega_\mathrm{r},\omega_\mathrm{nr})$ being real and normalized, we get $\left(\mathcal{P}_{l,m}\right)^{*} = \mathcal{P}_{-l,-m}$ and $\mathcal{P}_{0,0}(\omega_\mathrm{r},\omega_\mathrm{nr})  = 1/(4 \pi^2)$, respectively. Substituting the above expansion in the definition of the order parameters, we get
\begin{align}
     r^\mathrm{st}_\mathrm{r}e^{i\psi^\mathrm{st}_\mathrm{r}}  &=  4 \pi^2 \int_{-\infty}^\infty d \omega_\mathrm{r}d \omega_\mathrm{nr} g(\omega_\mathrm{r}) g(\omega_\mathrm{nr}) \mathcal{P}_{-1,0}, \label{order fourier sup1}\\
     r^\mathrm{st}_\mathrm{nr}e^{i\psi^\mathrm{st}_\mathrm{nr}} & =  4 \pi^2 \int_{-\infty}^\infty d \omega_\mathrm{r} d \omega_\mathrm{nr}g(\omega_\mathrm{r}) g(\omega_\mathrm{nr}) \mathcal{P}_{0,-1}. \label{order fourier sup2}
\end{align}
Further, substituting the Fourier expansion~\eqref{eq: Fourier Expansion} in Eq.~\eqref{eq: FP Kuramoto Finite Resetting} along with the steady state condition $\partial P/\partial t = 0$, we obtain
 \begin{align}
     &\left[(l^2+m^2)T+i(l \omega_\mathrm{r}+m\omega_\mathrm{nr})+\lambda\right] \mathcal{P}_{l,m} \nonumber \\
     &+\gamma\left(l\mathcal{P}_{l+1,m}+m\mathcal{P}_{l,m+1}\right)-\gamma^{*}\left(l\mathcal{P}_{l-1,m}+m\mathcal{P}_{l,m-1}\right) \nonumber \\
     & =  \lambda\left[ \alpha  + (-1)^l(1-\alpha) \right] \mathcal{P}_{0,m}, \label{eq: Fourier Relation Kuramoto SubReset Sup}
\end{align}  
with $\gamma \equiv K\left[ f r^\mathrm{st}_\mathrm{r}e^{i \psi^\mathrm{st}_\mathrm{r}}+\bar{f} r^\mathrm{st}_\mathrm{nr} e^{i \psi^\mathrm{st}_\mathrm{nr}} \right]/2$. From Eqs.~\eqref{order fourier sup1} and \eqref{order fourier sup2}, it is clear that our objects of interest are $\mathcal{P}_{-1,0}(\omega_\mathrm{r},\omega_\mathrm{nr})$ and $\mathcal{P}_{0,-1}(\omega_\mathrm{r},\omega_\mathrm{nr})$. Let us first focus on finding $\mathcal{P}_{-1,0}(\omega_\mathrm{r},\omega_\mathrm{nr})$. Putting $m=0$ in Eq.~\eqref{eq: Fourier Relation Kuramoto SubReset Sup}, we get
\begin{align}
    &\left[l^2T+il \omega_\mathrm{r}+\lambda\right] \mathcal{P}_{l,0}+l \gamma \mathcal{P}_{l+1,0}-l \gamma^{*}\mathcal{P}_{l-1,0}\nonumber\\
    &= \frac{\lambda}{4\pi^2}\left[ \alpha  + (-1)^l(1-\alpha) \right] 
  \label{eq: Fourier Relation l-axis Kuramoto SubReset Sup} .
\end{align}
Equation~\eqref{eq: Fourier Relation l-axis Kuramoto SubReset Sup} relates three consecutive $\mathcal{P}_{l,0}$'s in a linear relation. Among this string of $\mathcal{P}_{l,0}$'s with $l \geq 0$, we know the value at one end of the string, i.e., of the quantity $\mathcal{P}_{0,0}$. Using this, we may then express each $\mathcal{P}_{l,0}$ as a function of $\mathcal{P}_{l-1,0}$. Using the ansatz $\mathcal{P}_{l+1,0} = \Gamma_{l+1} \mathcal{P}_{l,0} + \Delta_{l+1}$~\cite{risken1980solutions} in Eq.~\eqref{eq: Fourier Relation l-axis Kuramoto SubReset Sup}, we get $\Gamma_{l}  = l \gamma^{*}/P_l$ and $\Delta_l = \{\lambda\left[ \alpha  + (-1)^l(1-\alpha) \right]/4\pi^2-l \gamma  \Delta_{l+1}\}/P_l$, where we have $P_l \equiv \left(l^2T+il \omega_\mathrm{r}+\lambda\right)+l \gamma \Gamma_{l+1}$. Putting $l=1$, we thus get $\mathcal{P}_{1,0} = \Gamma_{1} /(4\pi^2) + \Delta_{1}$, where both $\Gamma_1$ and $\Delta_1$ have continued fraction forms. For example, $\Gamma_1$ is given by
\begin{align}
    \Gamma_1(\omega_\mathrm{r}) &= \frac{ \gamma^{*}}{\left(T+i \omega_\mathrm{r}+\lambda\right)+ \gamma\left[ \frac{2 \gamma^{*}}{\left(4 T+ 2 i \omega_\mathrm{r}+\lambda\right)+2 \gamma \left[\ddots\right]}\right]}.
\end{align}
 Using the fact that $(\mathcal{P}_{-1,0}) = (\mathcal{P}_{1,0})^{*}$, and putting the expression of $\mathcal{P}_{-1,0}$ in Eq.~\eqref{order fourier sup1}, we obtain
 
\begin{align}
    r^\mathrm{st}_\mathrm{r}e^{i\psi^\mathrm{st}_\mathrm{r}} = 4 \pi^2 \int_{-\infty}^\infty d \omega_\mathrm{r} g(\omega_\mathrm{r}) \left[\frac{\Gamma^{*}_1 (\omega_\mathrm{r}) }{4 \pi^2} + \Delta^{*}_1(\omega_\mathrm{r}) \right].
\end{align}
Converting the integral into a contour integral in the lower-half of the complex-$\omega_{\mathrm{r}}$ plane and evaluating it using the residue theorem yield the self-consistent relation
\begin{align}
    r^\mathrm{st}_\mathrm{r}e^{i\psi^\mathrm{st}_\mathrm{r}} = 2 \pi i  \sum_{\omega_q } \mathrm{Res} \left.\left[g(\omega)\left(\Gamma^{*}_1 + 4 \pi^2\Delta^{*}_1 \right) \right]\right|_{\omega=\omega_q}, \label{eq: Finite Kuramoto SelfCon1 sup}
\end{align}
with $\omega_q$ the poles of $g(\omega)$ in the lower-half complex-$\omega_{\mathrm{r}}$ plane. 

Similarly, putting $l=0$ in Eq.~\eqref{eq: Fourier Relation Kuramoto SubReset Sup}, and following the same procedure as done in the case of $m=0$, we get that
\begin{align}
    r^\mathrm{st}_\mathrm{nr}e^{i\psi^\mathrm{st}_\mathrm{nr}} = 2 \pi i  \sum_{\omega_q } \mathrm{Res}\left. \left[g(\omega)~\Lambda^{*}_1\right]\right|_{\omega=\omega_q} \label{eq: Finite Kuramoto SelfCon2 sup},
\end{align}
where we have $\Lambda_m \equiv m\gamma^{*}/Q_m$ and $Q_m \equiv m^2T+im\omega_\mathrm{nr}+m \gamma \Lambda_{m+1}$. Equations~\eqref{eq: Finite Kuramoto SelfCon1 sup}~and~\eqref{eq: Finite Kuramoto SelfCon2 sup} are Eqs.~\eqref{eq: Finite Kuramoto SelfCon1} of the main text. Solving Eqs.~\eqref{eq: Finite Kuramoto SelfCon1 sup} and \eqref{eq: Finite Kuramoto SelfCon2 sup}
 simultaneously, we get the steady-state order parameters of the reset and non-reset subsystems. For numerically solving these equations, we truncate the continued fraction expressions of $\Gamma_1,~\Delta_1$ and $\Lambda_1$.\\

\textit{Appendix D: $\lambda \to \infty$ results for Kuramoto model--}
\begin{figure}[h!]
    \centering
\includegraphics[width=8.0cm]{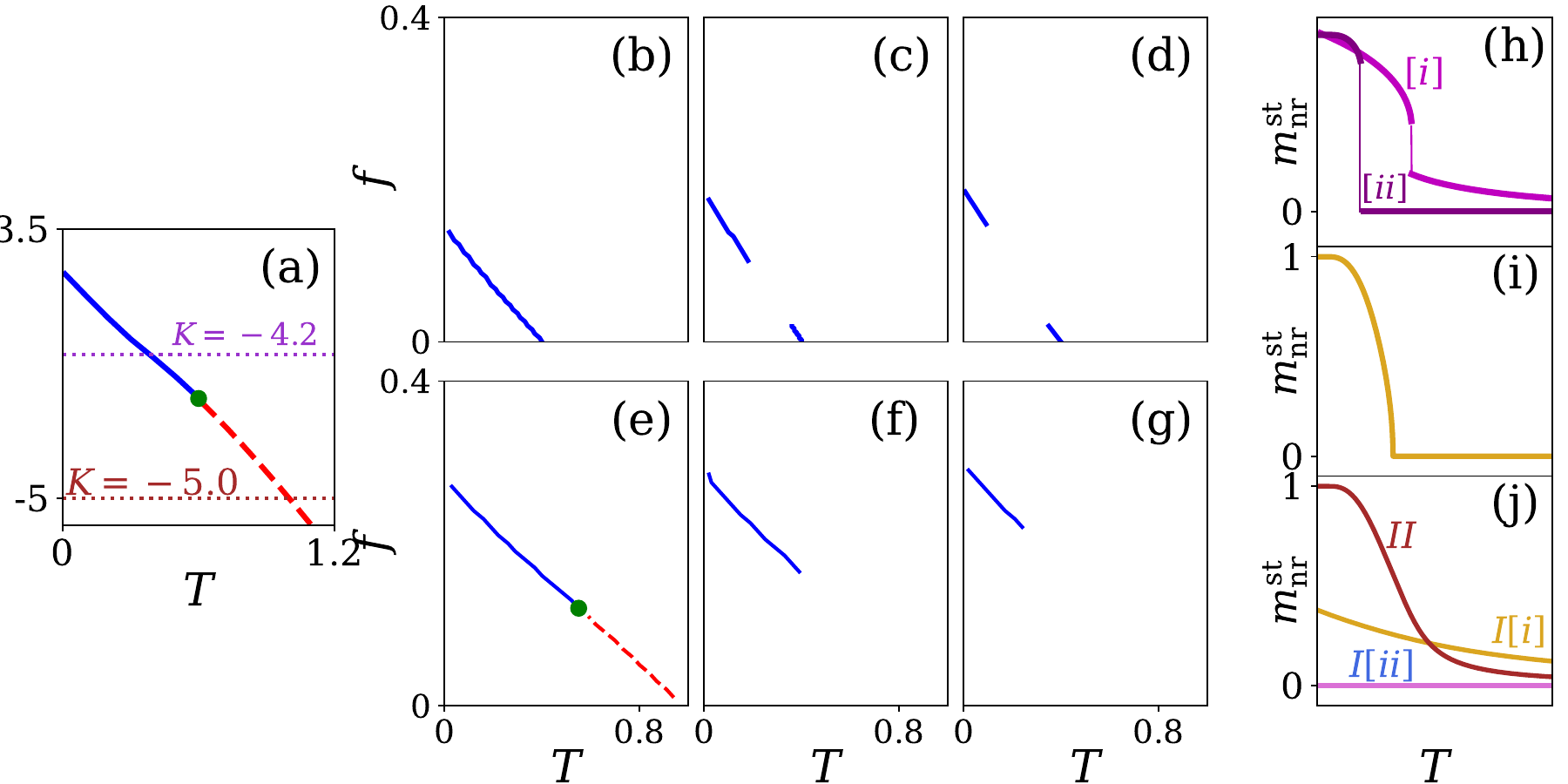}
    \caption{Phase diagram of Kuramoto model~\eqref{eq:BEG Hamiltonian} in (a) $(K,T)$-plane without resetting, (b) -- (g) $(f,T)$-plane with subsystem resetting to configurations with varying order $r_0$ at rate $\lambda \to \infty$ and $K = -4.2$ (b) -- (d), $K = -5.0$ (e) -- (g), using $\lambda\to \infty$ limit of Eq.~\eqref{eq: Finite Kuramoto SelfCon1}; $r_0=0,0.08,0.09$ (b) -- (d) and $r_0=0,0.006,0.02$ (e) -- (g). Schematic non-reset order parameter $r^\mathrm{st}_\mathrm{nr}$ vs. $T$ across (1) a first-order transition line in the phase diagrams: panel (h), with $[i]$ (respectively, $[ii]$) corresponding to $r_0 \neq 0$ (respectively, $r_0 = 0$);~(2) a continuous transition line in the phase diagrams: panel (i);~(3) a region without transition: panel (j), with $I$ and $II$ corresponding respectively to whether the region lies above or below a first-order-transition region: $I[i]$ (respectively, $I[ii]$) is for $r_0 \neq 0$ (respectively, $r_0 = 0$). 
}
\label{fig:7}
\end{figure}

\newpage
\textit{Appendix E: Finite-$\lambda$ results for Kuramoto model--}
\begin{figure}[H]
    \centering
\includegraphics[width=8.0cm]{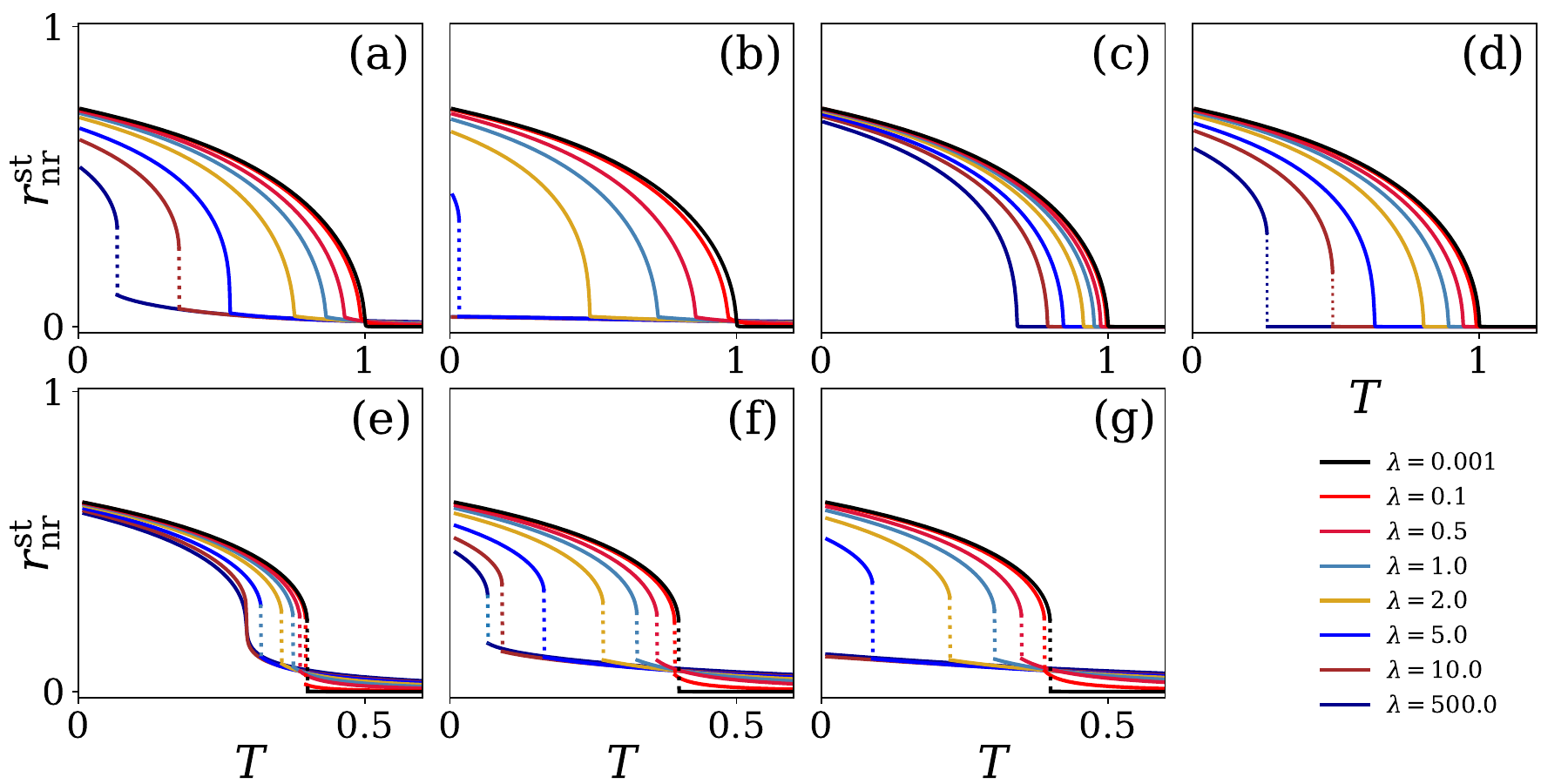}
    \caption{For the Kuramoto model~\eqref{eq: Kuramoto Dynamics} with $K=-5.0$~(a) -- (d) and $K=-4.2$~(e) -- (g), the figure shows $r^\mathrm{st}_\mathrm{nr}$ versus $T$, with increase of $\lambda$, from that of bare model ($\lambda \to 0$) to that in the limit $\lambda \to \infty$: $\lambda = 0.001, 0.1, 0.5, 1.0, 2.0, 5.0, 10.0, 500.0$, right to left in each panel. The data are obtained from solving Eqs.~\eqref{eq: Finite Kuramoto SelfCon1 sup}, \eqref{eq: Finite Kuramoto SelfCon2 sup}; results for $\lambda=500.0$ coincide with $\lambda\to\infty$ results. The panels correspond to different $f$ and $r_0$: (a),~(b): $f = 0.2, 0.5$, $r_0 = 0.02$; (c),~(d): $f = 0.09, 0.2$, $r_0 = 0$; (e),~(f),~(g): $f =0.05, 0.15, 0.2$, $r_0 = 0.08$.}\label{fig:8}
\end{figure}
\textit{Appendix F: Results for heterogeneous networks--}
\begin{figure}[H]
    \centering
\includegraphics[width=5. cm]{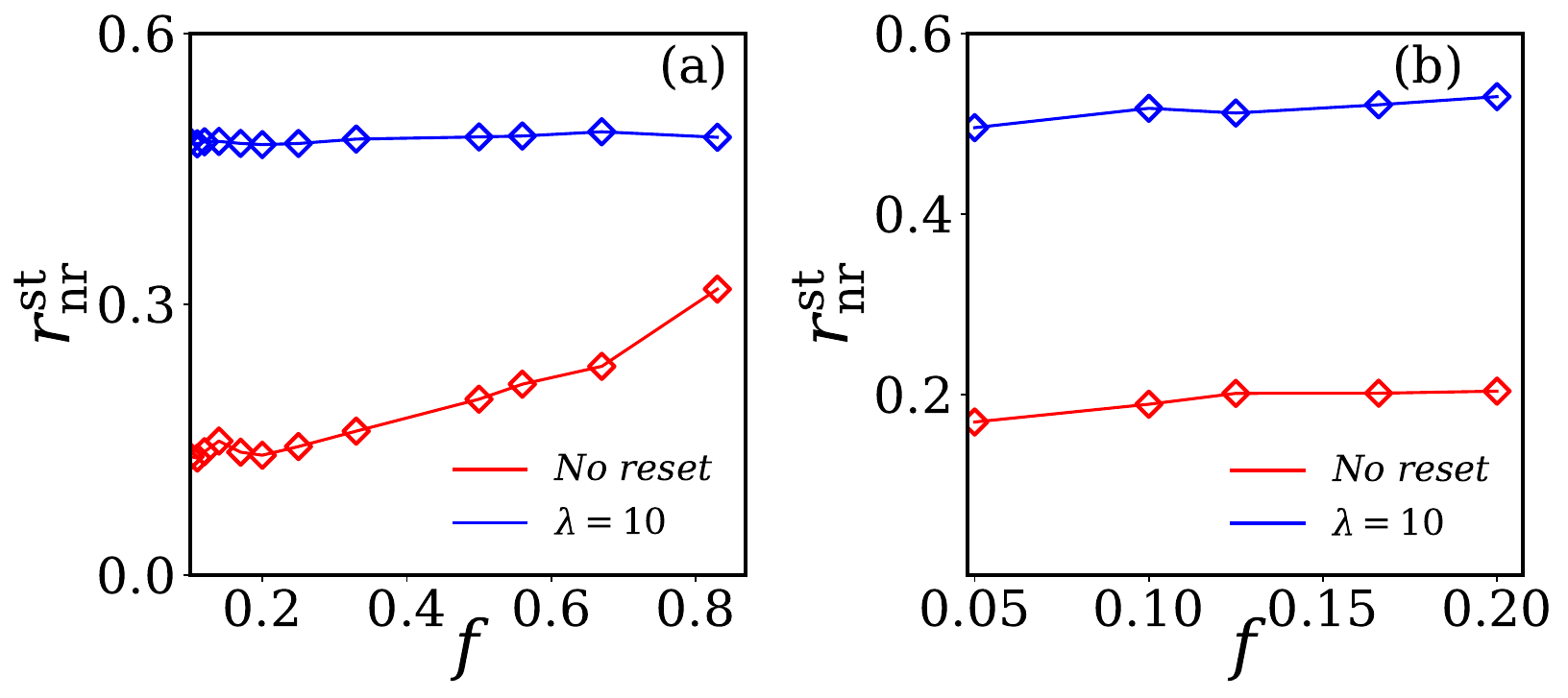}
    \caption{$r_\mathrm{nr}^{st}$ versus $f$ for Kuramoto oscillators on heterogeneous networks, in presence (blue line) and absence (red line) of resetting (panel (a) for model (i) and panel (b) for model (ii), see text). For (i), $n=100$, while $N$ varies from $120$ to $2000$. For (ii), $n=100$ highest-connectivity nodes, while $N$ varies from $500$ to $2000$. Measurable change in the synchronization behavior with respect to the bare dynamics is observed even as $f \to 0$.}\label{fig:9}
\end{figure}
\vspace{-0.3 cm}
We examine subsystem resetting in noiseless Kuramoto oscillators on two highly-heterogeneous networks: $(\mathrm{i})$ Inspired by Ref.~\cite{PhysRevLett.96.240602}, we consider a hybrid network of size $N$ comprising $n$ all-to-all connected hubs (denoted by $i=1,2,\ldots,n$) and $N-n$ non-hubs (denoted by $l=n+1,n+2,\ldots,N$) arranged as nearest-neighbors on a ring. Each hub connects to a non-hub with probability $1/2$. The bare evolution is given by $d \theta_i/dt = \omega_i + (K/n)\sum_{j=1}^n \sin {(\theta_j-\theta_i)} + [2 K/(N-n)]\sum_{l=n+1}^{N} A_{li} \sin {(\theta_l-\theta_i)},~d \theta_l/dt = \omega_l +(2K/n)\sum_{i=1}^n  A_{li} \sin {(\theta_i-\theta_l)}+ (K/2)\sin {(\theta_{l+1}-\theta_l)}+(K/2)\sin {(\theta_{l-1}-\theta_l)}$, with $A_{li}=1$ if a hub $i$ connects to a non-hub $l$, and $0$ otherwise. We reset only the hubs. On increasing $N$ at fixed $n$, $f = n/N \to 0$, and the average connectivity of the hubs diverges, while that of the non-hubs remains finite, making it a highly-heterogeneous network. $(\mathrm{ii})$ A scale-free network, where each oscillator evolves as~\cite{RODRIGUES20161} $d \theta_i/dt = \omega_i +[K/(\sum_j A_{ji})]\sum_{j} A_{ji} \sin {(\theta_j-\theta_i)}$, with $n_i \equiv \sum_j A_{ji}$ following a power-law distribution $p(n_i)\sim n_i^{-2}$. Here we reset the subset of nodes with highest connectivity. Results for the non-reset subsystem are shown in Fig.~\ref{fig:9}.

\newpage

\onecolumngrid

\setcounter{equation}{0}
\setcounter{section}{0}

\begin{center}
	{\bf Supplementary Information for: ``Manipulating phases in many-body interacting systems with subsystem resetting''}
\end{center}

\vspace{- 1.2 cm}
\section{Derivation of Eqs.~(4) and (5)}\label{sec:AppendixA}
Our starting point is Eq.~(1) of the main text describing the Hamiltonian of the BEG model.  As mentioned therein, $n$ out of the $N$ spins, indexed by $i = 1,2,\ldots, n$, are reset to a configuration $s_i = s_i^{(0)}; i = 1,2,\ldots,n$, with magnetization $m_0$ (clearly, $m_0 = (1/n) \sum_{i=1}^ns_i^{(0)}$) at rate $\lambda \to \infty$. As a result, the reset subsystem is frozen in the reset configuration. In this scenario, up to a constant shift, the effective Hamiltonian of the non-reset subsystem with configuration $\mathcal{C} \equiv \{s_{n+1}, s_{n+2}, \ldots, s_N \}$ becomes
\begin{align}
	H_{\mathrm{BEG}}^{\lambda \to \infty} = K\sum_{s_i \in \mathcal{C}} s_i^2 - f m_0 \sum_{s_i \in \mathcal{C}} s_i - \frac{1}{2N} \sum_{s_i,s_j \in \mathcal{C}} s_i s_j,\label{Eq:BEG_effective_Hamiltonian}
\end{align}
where the the energy scale have been shifted by a constant amount $\delta_\mathrm{E} = K \sum_{i=1}^n \left(s_i^{(0)}\right)^2 -  N f^2 m_0^2/2 $. Equation~\eqref{Eq:BEG_effective_Hamiltonian} is Eq.~$\mathrm{(4)}$ of the main text. The corresponding partition function of the non-reset subsystem becomes 
\begin{align}
	Z=\sum_{\mathcal{C}} e^{-\beta K\sum_{s_i \in \mathcal{C}} s_i^2 + \beta  f m_0\sum_{s_i \in \mathcal{C}} s_i + \frac{\beta}{2N} \sum_{s_i,s_j \in \mathcal{C}} s_i s_j}.\label{partion_function}
\end{align}

Using the Hubbard-Stratonovich (HS) transformation $\exp (b t^2)= \sqrt{b/\pi}\int_{-\infty}^{\infty} dx~\exp (- b x^2 + 2 t b x);~b>0$, we write Eq.~\eqref{partion_function} as
\begin{align}
	Z&= \sqrt{\frac{\beta  N}{2 \pi}} \bar{f} \sum_{\mathcal{C} } e^{-\beta K\sum_{s_i \in \mathcal{C}} s_i^2 + \beta f  m_0\sum_{s_i \in \mathcal{C}} s_i} \int_{-\infty}^{\infty} dx~e^{-\frac{\beta \bar{f} x^2}{2}  +\beta N \bar{f}^2 \sum_{s_i \in \mathcal{C}} s_i x }\nonumber\\
	&=\sqrt{\frac{\beta N}{2 \pi}} \bar{f}\int_{-\infty}^{\infty} dx~\left(\sum_{s_i = 0, \pm1 } e^{-\beta K  s_i^2 + \beta f m_0 s_i +\beta  \bar{f}  s_i x }\right)^{N-n} e^{-\frac{\beta}{2} N \bar{f}^2 x^2 }\nonumber\\
	&=\sqrt{\frac{\beta N}{2 \pi}} \bar{f}\int_{-\infty}^{\infty} dx~e^{-N \beta \widetilde{F}(\beta,x,f)} , \label{Eq:BEG_saddle_integral}
\end{align}
with
\begin{align} 
	\widetilde{F}(\beta,m_\mathrm{nr},f)=\frac{\bar{f}^2m_\mathrm{nr}^2}{2}-\frac{\bar{f}}{\beta} \ln \left\{1+ 2 e^{-\beta K}\cosh[\beta  (fm_0 +\bar{f}m_\mathrm{nr})]\right\}.
\end{align}
As detailed in the text, evaluating the integral in Eq.~\eqref{Eq:BEG_saddle_integral} by the saddle-point method yields free-energy/spin as $\tilde{F}(\beta,m^\mathrm{st}_{\mathrm{nr}},f)$, with steady-state magnetization $m^\mathrm{st}_{\mathrm{nr}}$ satisfying the self-consistent relation given in Eq. $\mathrm{(5)}$ of the main text.
\section{Derivation of Glauber Dynamics of BEG Model}
Defining the rate of transition from configuration $\{s_i\}$ to $\{s'_i \}$ as $w(s_i \rightarrow s_i')$, the time dependence of the ensemble average of an observable, namely, the value $s_i$ of the $i$-th spin, may be written as~\cite{10.1063/1.1703954}
\begin{align}
	\frac{d \langle s_i \rangle}{d t } = \left \langle \sum_{s'}(s'_i-s_i) w(s_i \to s'_i) \right \rangle .\label{Eq:time_dependence_Glauber}
\end{align}
From the condition of detailed balance, we have
\begin{align}
	w(s_i \to s'_i) = \frac{e^{-\beta\Delta H (s_i \to s'_i)}}{\sum_{s'}e^{-\beta\Delta H (s_i \to s'_i)}},
\end{align}
where $\Delta H (s_i \to s'_i)$ is the change in the energy of the system due to the flipping of the $i$-th spin from the value $s_i$ to $s'_i$. Defining $z = \frac{1}{N}\sum_{\substack{i=1}}^N s_i$, we calculate the transition rate between several configurations and obtain $w(+1 \to -1) = \frac{e^{- \beta  J z }}{2 \cosh{\beta J z} + e^{\beta K} }$, $w(+1 \to 0)=\frac{e^{\beta K}}{2 \cosh{\beta J z} + e^{\beta K} }$, $w(0 \to +1) = \frac{e^{ \beta  J z }}{2 \cosh{\beta J z} + e^{\beta K} }$, $w(0 \to -1) = \frac{e^{ - \beta  J z }}{2 \cosh{\beta J z} + e^{\beta K} }$, $w(-1 \to +1) = \frac{e^{ \beta  J z }}{2 \cosh{\beta J z} + e^{\beta K}}$,
and $w(-1 \to 0) 
= \frac{e^{ \beta  K }}{2 \cosh{\beta J z} + e^{\beta K} }$. Using these in Eq.~\eqref{Eq:time_dependence_Glauber}, we obtain
\begin{align}
	\frac{d \langle s_i \rangle}{d t } 
	&= - \left \langle s_i \sum_{s'} w(s_i \to s'_i) \right \rangle + \left \langle \sum_{s'}s'_i w(s_i \to s'_i) \right \rangle \nonumber\\
	&= - \left \langle s_i \right \rangle + \left \langle \sum_{s'}s'_i w(s_i \to s'_i) \right \rangle = - \left \langle s_i \right \rangle + \left \langle  \frac{ 2 \sinh{\beta J z} }{2 \cosh{\beta J z} + e^{\beta K} }\right \rangle \label{Eq:dsk/dt11}.
\end{align}

Further, we have 
\begin{align}
	z &= \frac{1}{N}\sum_{\substack{j=1}}^N s_i =\frac{1}{N}\sum_{i=1}^N (\langle s_i \rangle+\delta_i)= m + \delta,
\end{align}
with $~\delta_i \equiv s_i - \langle s_i\rangle$ and $\delta \equiv \frac{1}{N}\sum_{i=1}^N \delta_i$. Being a mean-field system, we have $\langle \delta_i \rangle = 0~\forall~i$. As a result, $\langle z\rangle=m$, and the fluctuations $\delta$ must satisfy
\begin{align}
	\langle \delta \rangle = \frac{1}{N}\sum_{i=1}^N \langle \delta_i \rangle = 0.
\end{align}
Furthermore, being mean-field allows to write $\langle s_i s_j \rangle=\langle s_i \rangle \langle s_j\rangle$, implying that $\langle \delta_i \delta_j\rangle=\langle \delta_i \rangle \langle \delta_j\rangle=0$. In fact, one has $\langle s_i s_j s_k \ldots \rangle=\langle s_i \rangle \langle s_j \rangle \langle s_k \rangle \ldots$, implying $\langle \delta_i \delta_j \delta_k \ldots\rangle=\langle \delta_i\rangle \langle \delta_j \rangle \langle \delta_k \rangle \ldots=0$. We then get
\begin{align}
	\langle \delta^k \rangle = \frac{1}{N^k} \sum_{i_1} \cdots \sum_{i_k} \langle \delta_{i_1} \cdots \delta_{i_k} \rangle = 0.
\end{align}
At this point, putting $z = m + \delta$ in Eq.~\eqref{Eq:dsk/dt11}, expanding the second term on the right hand side as a power of $\delta$ and using $\langle \delta^k  \rangle =  0~\forall~k$, we get
\begin{align}
	\frac{d \langle s_i \rangle}{d t } = - \left \langle s_i \right \rangle + \frac{ 2 \sinh{\beta J m} }{2 \cosh{\beta J m} + e^{\beta K} } \label{Eq:dsk/dt1}.
\end{align}

We now define the order parameters of the reset and non-reset subsystems as $m_{\mathrm{r}} = \left \langle \frac{1}{n} \sum_{i = 1}^n s_i \right \rangle$ and $m_{\mathrm{nr}} = \left \langle \frac{1}{N-n} \sum_{i = n+1}^N s_i \right \rangle$, respectively. Putting them back into Eq.~\eqref{Eq:dsk/dt1}, summing the equation for $ i = 1,2,\ldots, n$ and dividing by $n$, we get the evolution equation of the order parameter of the reset subsystem, which reads as
\begin{align}\label{sup-Glauber1}
	\frac{d m_\mathrm{r}}{d t }  = - m_\mathrm{r} +    \frac{ 2 \sinh{[\beta (f m_\mathrm{r} + \bar{f} m_\mathrm{nr})]} }{2 \cosh{[\beta (f m_\mathrm{r} + \bar{f}m_\mathrm{nr})]} + e^{\beta K} } .
\end{align}
Similarly, summing both sides of Eq.~\eqref{Eq:dsk/dt1} for $ i = n+1,\ldots, N$ and divining by $(N-n)$, we get the evolution equation of the order parameter of the non-reset subsystem, which reads as
\begin{align}\label{sup-Glauber2}
	\frac{d m_\mathrm{nr}}{d t }  = - m_\mathrm{nr} +   \frac{ 2 \sinh{[\beta (f m_\mathrm{r} + \bar{f} m_\mathrm{nr})]} }{2 \cosh{[\beta (f m_\mathrm{r} + \bar{f}m_\mathrm{nr})]} + e^{\beta K} } .
\end{align}
Equations~\eqref{sup-Glauber1} and \eqref{sup-Glauber2} are quoted in the main text.

\section{Derivation of Eqs.~(7) and (8)}
Similar to the BEG model, here also our starting point is Eq.~(2) of the main text describing the Hamiltonian of the KN model. Here also, if spins with index $i = 1, 2, \ldots, n$ are reset to a configuration $s_i = s_i^{(0)}; i = 1,2,\ldots,n$, with magnetization $m_0$ (clearly, $m_0 = (1/n) \sum_{i=1}^ns_i^{(0)}$) at rate $\lambda \to \infty$, we may write the effective Hamiltonian of the non-reset subsystem with configuration $\mathcal{C} \equiv \{s_{n+1}, s_{n+2}, \ldots, s_N \}$ as
\begin{align}
	H_{\mathrm{KN}}^{\lambda \to \infty} 
	=\frac{K}{2}\sum_{s_i \in \mathcal{C}} (s_i s_{i+1}-1)- f m_0 \sum_{s_i \in \mathcal{C}} s_i  -\frac{1}{2 N} \sum_{s_i \in \mathcal{C}} s_i s_j -\frac{K}{2} (s_N-1)(s_{n+1}-1) ,\label{Eq:KN_effective_Hamiltonian sup}
\end{align}
where the the energy scale have been shifted by a constant amount $\delta_\mathrm{E} = (K/2) \sum_{i=1}^{n-1} s_i^{(0)}s_{i+1}^{(0)} -  N f^2 m_0^2/2 $. For simplicity, here we have chosen $s_1^{(0)} = s_n^{(0)} = 1$. Equation.~\eqref{Eq:KN_effective_Hamiltonian sup} is Eq.~(7) of the main text. The corresponding partition function of the non-reset subsystem becomes 
\begin{align}
	Z = \sum_{\mathcal{C}} e^{\Bigl[\frac{\beta}{2N} \sum \limits_{s_i,s_j \in \mathcal{C}} s_i s_j + \beta  f m_0\sum \limits_{s_i \in \mathcal{C}}s_i - \frac{\beta K}{2}\sum \limits_{s_i \in \mathcal{C}} (s_i s_{i+1}-1) +\frac{\beta K}{2} (s_N-1)(s_{n+1}-1) \Bigl]} .\label{eq:Parttion_fucntion_KN}
\end{align} 

Using the $\mathrm{HS}$ transformation in Eq.~\eqref{eq:Parttion_fucntion_KN}, we obtain
\begin{align}
	Z(\beta, N-n) &= \sum_{\mathcal{C}} \sqrt{\frac{\beta N}{2 \pi }}\bar{f} \int_{-\infty}^{\infty} dx~e^{- \frac{\beta N \bar{f}^2}{2} x^2 +  \beta x \bar{f} \sum \limits_{s_i \in \mathcal{C}} s_i } e^{\beta f \sum \limits_{s_i \in \mathcal{C}} s_i - \frac{\beta K}{2}\sum \limits_{s_i \in \mathcal{C}} (s_i s_{i+1}-1)  +\frac{\beta K}{2} (s_N-1)(s_{n+1}-1) } ,\nonumber \\
	&= \sqrt{\frac{\beta N}{2 \pi }}\bar{f} \int_{-\infty}^{\infty} dx~e^{- \frac{\beta N (1-f)^2}{2} x^2} Z_0(\beta,x,f,N), \label{HS KN SUP}
\end{align}
where $Z_0(\beta, x, f, N)=\sum \limits_{\mathcal{C}} \exp \left(\beta M \sum \limits_{s_i \in \mathcal{C}} s_i - \frac{\beta K}{2}\sum \limits_{s_i \in \mathcal{C}} (s_i s_{i+1}-1)  +\frac{\beta K}{2} (s_N-1)(s_{n+1}-1) \right)$ is the partition function of the nearest neighbor Ising model in presence of an external field of strength $M \equiv f m_0+\bar{f} x$ and with two additional boundary terms. The quantity $Z_0(\beta, x, f, N)$ can be further simplified using the transfer matrices as follows:
\begin{align}
	Z_0(\beta,x,f,N) &= \sum_{\mathcal{C}}e^{\beta \left[M \sum \limits_{s_i \in \mathcal{C}} s_i - \frac{K}{2}\sum \limits_{s_i \in \mathcal{C}} (s_i s_{i+1}-1)  +\frac{K}{2} (s_N-1)(s_{n+1}-1) \right] }, \nonumber\\
	&= e^{-\frac{\beta K (N-n)}{2}} \sum_{s_{n+1} = -1}^{+1} \cdots \sum_{s_{N} = -1}^{+1} T_{s_{n+1}s_{n+2}} \cdots T_{s_{N-1}s_{N}} A_{s_{N}s_{n+1}},\label{eq:Z0_exp1}
\end{align}
where $T$ and $A$ are the transfer matrices with matrix elements
\begin{align}
	&T_{s_is_{i+1}} \equiv  e^{\beta \left( \frac{M}{2} s_i + \frac{M}{2}  s_{i+1} - \frac{K}{2} s_i s_{i+1} \right) },\\
	&A_{s_Ns_{n+1}} \equiv e^{\beta \left( \frac{M}{2} s_N + \frac{M}{2} s_{n+1} - \frac{K}{2} s_N s_{n+1} +\frac{K}{2} (s_N-1)(s_{n+1}-1) \right) }.
\end{align}
Since $s_i$ can take values $\pm 1$, we can write $T$ and $A$ as follows:
\begin{align}
	T = \left[ \begin{array}{cc}
		e^{\beta \left( M - \frac{K}{2} \right) } & e^{\frac{\beta K}{2}} \\
		e^{\frac{\beta K}{2}} & e^{-\beta \left( M + \frac{K}{2} \right) } 
	\end{array}
	\right];~~ A = \left[ \begin{array}{cc}
		e^{\beta \left( M - \frac{K}{2} \right) } & e^{\frac{\beta K}{2}} \\
		e^{\frac{\beta K}{2}} & e^{-\beta \left( M -  \frac{3 K}{2} \right) }
	\end{array}
	\right], 
\end{align}
which can be used to rewrite Eq.~\eqref{eq:Z0_exp1}  as $ Z_0=e^{-\frac{\beta K (N-n)}{2}} \mathrm{Tr} \left[ T^{N-n-1}A \right]$. 

We now go in the eigenbasis of $T$.  From the expression of $T$, we straightforwardly get the eigenvalues $\lambda$ as
\begin{align}
	\lambda_{\pm} = e^{-\frac{\beta K}{2}} \left[ \cosh{\beta M} \pm \sqrt{ \cosh^2{\beta M} + 2 e^{\beta K} \sinh{\beta K} }\right],
\end{align}
and the matrix that diagonalizes $T$ as
\begin{align}
	C = \left[ \begin{array}{cc}
		\frac{1}{\sqrt{1+b^2 (\lambda_{+}-ab)^2}} &  \frac{1}{\sqrt{1+b^2 (\lambda_{-}-ab)^2}}\\
		\frac{b (\lambda_{+}-ab)}{\sqrt{1+b^2 (\lambda_{+}-ab)^2}}   & \frac{b (\lambda_{-}-ab)}{\sqrt{1+b^2 (\lambda_{-}-ab)^2}}
	\end{array}
	\right],
\end{align}
with $a \equiv e^{\beta F}$ and $b \equiv e^{-\frac{\beta K}{2}}$. Then, we get
\begin{align}
	Z_0(\beta,x,f,N) = e^{-\frac{\beta K (N-n)}{2}} \mathrm{Tr} \left[ T^{N-n-1}A \right] = e^{-\frac{\beta K (N-n)}{2}} \mathrm{Tr} \left[ D^{N-n-1} \underbrace{C^{-1} A C}_{\tilde{A}}\right]= e^{-\frac{\beta K (N-n)}{2}} \mathrm{Tr} \left[ D^{N-n-1} \underbrace{C^{-1} A C}_{\tilde{A}}\right],
\end{align}
where we have
\begin{align}
	D^{N-n-1} = \left[ \begin{array}{cc}
		\lambda^{N-n-1}_{+} &  0\\
		0  & \lambda^{N-n-1}_{-}
	\end{array}
	\right].
\end{align}

Upon further computation, we get that
\begin{align}
	Z_0(\beta,x,f,N) = e^{-\frac{\beta K (N-n)}{2}} \left( \lambda^{N-n-1}_{+} \tilde{A}_{11} +  \lambda^{N-n-1}_{-} \tilde{A}_{22}\right) \approx e^{-\frac{\beta K (N-n)}{2}}  \lambda^{N-n-1}_{+} \tilde{A}_{11},
\end{align}
where we have used the fact that as $\lambda_{+} > \lambda_{-}$, in the thermodynamic limit, one has $\lambda^{N-n-1}_{+} \gg \lambda^{N-n-1}_{-}$. Putting all of these together in Eq.~\eqref{HS KN SUP}, we finally get
\begin{align}
	Z(\beta, N-n) = \sqrt{\frac{\beta  N}{2 \pi}} \bar{f}\int_{-\infty}^{\infty} dx~e^{-N \beta \tilde{F}(\beta,x,f)},
\end{align}
with
\begin{align}
	\tilde{F}(\beta,x,f)
	&= \frac{\bar{f}^2}{2} x^2  - \frac{\bar{f}}{\beta} \log  \left[ \cosh{\beta M} + \sqrt{ \cosh^2{\beta M} + 2 e^{\beta K} \sinh{\beta K} }\right] + K \bar{f} .
\end{align}
Following similar arguments as in Sec.~(\ref{sec:AppendixA}), the steady-state magnetization $m^\mathrm{st}_{\mathrm{nr}}$ satisfies the self-consistent Eq.~$\mathrm{(8)}$ of main text.

\section{$\lambda \to \infty$ limit of Eqs.~(10)}

We start from Eq. (14) in Appendix C of the main text, which reads as
\begin{align}
	&\left[(l^2+m^2)T+i(l \omega_\mathrm{r}+m\omega_\mathrm{nr})+\lambda\right] \mathcal{P}_{l,m} +\gamma\left(l\mathcal{P}_{l+1,m}+m\mathcal{P}_{l,m+1}\right)-\gamma^{*}\left(l\mathcal{P}_{l-1,m}+m\mathcal{P}_{l,m-1}\right) \nonumber \\
	& =  \lambda\left[ \alpha  + (-1)^l(1-\alpha) \right] \mathcal{P}_{0,m}. \label{eq: Fourier Relation Kuramoto SubReset Sup}
\end{align}  
Putting $m = 0$ and $l=0$ respectively in Eq.~\eqref{eq: Fourier Relation Kuramoto SubReset Sup}, we get
\begin{align}
	&\left[l^2T+il \omega_\mathrm{r}+\lambda\right] \mathcal{P}_{l,0} +\gamma l\mathcal{P}_{l+1,0}-\gamma^{*}l\mathcal{P}_{l-1,0}  =  \frac{\lambda}{4\pi^2}\left[ \alpha  + (-1)^l(1-\alpha) \right] , \label{m = 0 eq}\\
	&\left[m^2T+i m\omega_\mathrm{nr}\right] \mathcal{P}_{0,m} +\gamma m\mathcal{P}_{0,m+1}-\gamma^{*}m\mathcal{P}_{0,m-1}  = 0 .\label{l = 0 eq}
\end{align}

Now, $\mathcal{P}_{l,m}$'s are finite for all $l,m$, as the probability density $P(\theta_\mathrm{r}, \theta_\mathrm{nr}, \omega_\mathrm{r}, \omega_\mathrm{nr}, t)$ is a finite quantity. As a result, putting $\lambda \to \infty$ in Eq.~\eqref{m = 0 eq}, we get
\begin{align}
	\mathcal{P}_{l,0}   =  \frac{1}{4\pi^2}\left[ \alpha  + (-1)^l(1-\alpha) \right] . \label{ m = 0 eq lam infinite limit}
\end{align}  
Putting it in the definition of $r^\mathrm{st}_\mathrm{nr}$, we get
\begin{align}
	r^\mathrm{st}_\mathrm{r} e^{i\psi^\mathrm{st}_\mathrm{r}} = 4 \pi^2 \int_{-\infty}^\infty d \omega_\mathrm{r}d \omega_\mathrm{nr} g(\omega_\mathrm{r}) g(\omega_\mathrm{nr}) \mathcal{P}_{-1,0} = 2\alpha -1,
\end{align}
which implies that $r^\mathrm{st}_\mathrm{r} = |2\alpha -1| = r_0$. This we can also physically understand from the fact that in the limit $\lambda \to \infty$, the reset subsystem gets frozen in the reset configuration. Now, putting this in the expression for the quantity $\gamma$ as defined in the Appendix C of the main text, we get
\begin{align}
	\gamma_\infty = \frac{K}{2}\left[ f r_0+\bar{f} r^\mathrm{st}_\mathrm{nr} e^{i \psi^\mathrm{st}_\mathrm{nr}} \right],
\end{align}
which is a function of only $r^\mathrm{st}_\mathrm{nr}$ and $\psi^\mathrm{st}_\mathrm{nr}$. 

Solving Eq.~\eqref{l = 0 eq} in the same method as discussed in Appendix C, we get
\begin{align}
	\mathcal{P}_{0,1} = \frac{\Lambda^{(\infty)}_1(\omega_\mathrm{nr})}{4\pi^2},
\end{align}
where we have
\begin{align}
	\Lambda^{(\infty)}_1 (\omega_\mathrm{nr}) = \frac{ \gamma_\infty^{*}}{\left(T+i \omega_\mathrm{nr}\right)+ \gamma_\infty\left[ \frac{2 \gamma_\infty^{*}}{\left(4T+i2 \omega_\mathrm{nr}\right)+2 \gamma_\infty \left[\frac{3 \gamma_\infty^{*}}{\left(9 T+i3 \omega_\mathrm{nr}\right)+3 \gamma_\infty \left[\ddots \right]}\right]}\right]}.
\end{align}
Putting this expression in the definition of $r^\mathrm{st}_\mathrm{nr}$, we get
\begin{align}
	r^{\mathrm{st}}_\mathrm{nr}  e^{i \psi^{\mathrm{st}}_\mathrm{nr}}&= 2 \pi i  \sum_{\omega_q} \mathrm{Res}\left. \left\{g(\omega)\left[\Lambda^{(\infty)}_1 (\omega_\mathrm{nr})\right]^{*}\right\}\right|_{\omega=\omega_q},
\end{align}
where $\omega_q$'s are the poles of $g(\omega)$ in the lower-half of the  complex-$\omega$ plane.

\end{document}